\documentclass{article}
\PassOptionsToPackage{authoryear,round,semicolon}{natbib}
\usepackage[preprint]{iaseai26}

\usepackage{graphicx} 
\usepackage[utf8]{inputenc} 
\usepackage[T1]{fontenc}    
\usepackage{hyperref}       
\usepackage{url}            
\usepackage{booktabs}       
\usepackage{amsfonts}       
\usepackage{nicefrac}       
\usepackage{microtype}      
\usepackage{xcolor}
\usepackage{microtype}
\usepackage{hyperref}
\usepackage{booktabs}
\usepackage{tabularx}
\usepackage{enumitem}
\usepackage{array}
\usepackage{caption}
\usepackage{natbib}
\usepackage{multirow}
\usepackage{pdflscape} 
\usepackage{tikz}
\usetikzlibrary{positioning, shapes.geometric, arrows.meta}
\usepackage{amssymb} 
\usepackage{float} 
\usepackage{graphicx}
\usepackage{adjustbox}
 
\setlist[itemize]{nosep,leftmargin=*}
\usepackage[table]{xcolor}
\definecolor{HighRed}{HTML}{F9D5D3}
\definecolor{MedYellow}{HTML}{FFF7C2}
\definecolor{LowGreen}{HTML}{D8F2D0}

\title{How frontier AI companies could implement an internal audit function}

\author{
  Francesca Gomez$^{1}$\thanks{Corresponding author: francesca@arcadiaimpact.org}, Adam Buick$^{1}$$^{,2}$, Leah Ferentinos$^{1}$, Haelee Kim$^{1}$,
  Elley Lee$^{1*}$\\[0.5em]
  \small\textit{$^{1}$Arcadia Impact - AI Governance Taskforce, London, United Kingdom}\\
   \small\textit{$^{2}$Ulster University, Northern Ireland}
}

\date{December 2025}

\begin{document}

\maketitle

\begin{abstract}
Frontier AI developers operate at the intersection of rapid technical progress, extreme risk exposure, and growing regulatory scrutiny. While a range of external evaluations and safety frameworks have emerged, comparatively little attention has been paid to how internal organizational assurance should be structured to provide sustained, evidence-based oversight of catastrophic and systemic risks. This paper examines how an internal audit function could be designed to provide meaningful assurance for frontier AI developers, and the practical trade-offs that shape its effectiveness. Drawing on professional internal auditing standards, risk-based assurance theory, and emerging frontier-AI governance literature, we analyze four core design dimensions: (i) audit scope across model-level, system-level, and governance-level controls; (ii) sourcing arrangements (in-house, co-sourced, and outsourced); (iii) audit frequency and cadence; and (iv) access to sensitive information required for credible assurance. For each dimension, we define the relevant option space, assess benefits and limitations, and identify key organizational and security trade-offs.

We show that while model-level audits offer the most direct insight into dangerous capabilities, system-level and governance-level audits provide broader and more durable coverage as frontier risks increasingly depend on organizational controls rather than isolated model failures. We further demonstrate that hybrid sourcing models anchored by an internal Chief Audit Executive allow greater flexibility for tiered information access while preserving independence and external credibility. Finally, we argue that differentiated audit frequencies and carefully governed information access regimes are necessary to sustain assurance value in environments where both technical capabilities and organizational structures evolve rapidly.

Our findings suggest that internal audit, if deliberately designed for the frontier AI context, can play a central role in strengthening safety governance, complementing external evaluations, and providing boards and regulators with higher-confidence, system-wide assurance over catastrophic risk controls.
\end{abstract}

\clearpage
\section*{Executive summary}

This paper sets out the principal design choices for frontier AI developers implementing internal audit for assurance and addresses the practical questions of how to operationalize it.

\textbf{What is internal audit?} 
Internal audit is a management function designed to provide independent assurance over an organization’s governance, risk-management, and internal control processes by reporting directly to the board, thereby supporting both independence and direct access to senior management. Through its organizational design, it is intended to have unrestricted access to internal information in order to enable deeper and more continuous assurance than most external mechanisms, although in practice it may be fully or partially outsourced. It is now standard governance infrastructure across publicly listed firms, financial-services organizations, and public-sector institutions.

\textbf{Relevance to the EU AI Act.}
The EU General-Purpose AI Code of Practice (commitment 8) requires signatories to allocate responsibility for providing assurance about the adequacy of systemic risk management processes to the board or equivalent governance body. This analysis aims to inform implementation guidance for commitment 8 by examining how internal audit could fulfill this assurance function, and the practical trade-offs that shape internal audit effectiveness in frontier AI contexts.

\textbf{Practical design choices for implementing an internal audit function.} We identify four key aspects relevant to frontier AI developers implementing an internal audit function and evaluate options for each of them.

\begingroup
\setlength{\tabcolsep}{10pt}        
\renewcommand{\arraystretch}{1.3}   
\hyphenpenalty=10000
\exhyphenpenalty=10000
\begin{table}[ht]
\centering
\begin{tabular}{@{}p{0.40\textwidth} p{0.55\textwidth}@{}}
\toprule
\textbf{Question} & \textbf{Options} \\
\midrule
\begin{minipage}[t]{\linewidth}\raggedright
What risks should internal audit provide assurance about?
\end{minipage}
&
\begin{minipage}[t]{\linewidth}\raggedright
\begin{itemize}
    \item Model-level risks (e.g., dangerous capabilities, evaluation validity, post-training safeguards, refusal systems).
    \item System-level risks (e.g., access controls, logging, monitoring, incident response, secure compute, weight protection).
    \item Governance-level risks (e.g., board oversight, safety committees, release gates, escalation, override controls).
\end{itemize}
\end{minipage}
\\
\addlinespace[1em]
\begin{minipage}[t]{\linewidth}\raggedright
How should the internal audit function be sourced?
\end{minipage}
&
\begin{minipage}[t]{\linewidth}\raggedright
\begin{itemize}
    \item Fully in-house internal audit team.
    \item Co-sourced model (internal Chief Audit Executive + external specialists).
    \item Fully outsourced to an external provider.
\end{itemize}
\end{minipage}
\\
\addlinespace[1em]
\begin{minipage}[t]{\linewidth}\raggedright
How frequently should internal audits be conducted?
\end{minipage}
&
\begin{minipage}[t]{\linewidth}\raggedright
\begin{itemize}
    \item Annual cycles.
    \item Semi-annual or quarterly cycles.
    \item Ad-hoc reviews / rapid assurance.
    \item Continuous auditing.
\end{itemize}
\end{minipage}
\\
\addlinespace[1em]
\begin{minipage}[t]{\linewidth}\raggedright
What information sources should internal auditors access?
\end{minipage}
&
\begin{minipage}[t]{\linewidth}\raggedright
\begin{itemize}
    \item Structural information (e.g., governance frameworks, safety policies, organisation charts).
    \item Procedural information (e.g., internal reports, system documentation, control designs).
    \item Operational information (e.g., staff interviews, incident records, internal testing outputs).
    \item Technical information (e.g., model evaluations, logs, monitoring systems, codebases, infrastructure).
\end{itemize}
\end{minipage}
\\
\bottomrule
\end{tabular}
\end{table}
\endgroup

\textbf{What are the main implementation trade-offs?}
Design choices across these areas must balance competing priorities. Access to technical information enables higher-quality assurance but increases information-security risk; in-house or co-sourced models reduce security exposure but may face greater independence concerns than fully outsourced approaches; more frequent audits maintain assurance relevance but impose operational friction on development and safety teams. Importantly, organizations can customize their approaches over time to match the capability levels of their models, risk appetite, and security posture, building toward more comprehensive approaches over time by developing and testing solutions to address the challenges identified in this analysis.
\clearpage

\section*{Contribution Statement}

\noindent
\textbf{Conceptualization:} Francesca Gomez\\
\textbf{Methodology, Investigation:} All authors

\medskip

\noindent
\textbf{Writing -- Original Draft, Review:}

\begin{tabular}{@{}ll@{}}
Leah Ferentinos   & Section 2 \\
Elley Lee         & Section 3 \\
Adam Buick        & Section 4 \\
Haelee Kim        & Section 5 \\
\end{tabular}

\medskip

\noindent
\textbf{Writing -- Review \& Editing, Supervision:} Francesca Gomez (Lead Editor)

\medskip

\noindent
\textbf{Project Administration:} Francesca Gomez, Ben R.\ Smith

\section*{Acknowledgements}
We gratefully acknowledge Aidan Homewood for advice during the development of this paper, including feedback as we shaped the research question and comments that helped strengthen the clarity and accuracy of several sections. Any remaining errors are solely the authors'. We also wish to thank to Ben R. Smith of Arcadia Impact for his overarching management of the program under which this work was undertaken. 

\clearpage
\tableofcontents

\clearpage
\section{Introduction}

As the capabilities of frontier AI models advance \citep{epochaiDataAIBenchmarking2024}, so too does the systemic risk associated with their development and deployment, spanning from potential misuse by malicious actors to broader socioeconomic harms \citep{MITAIRisk}. Frontier model releases recently crossed new capability thresholds for chemical, biological, radiological, and nuclear (CBRN) weapons development \citep{openai_gpt_4_system_card_2023}; \citep{anthropic_claude_4_system_card_2025}, triggering stronger deployment and security protections under developers’ internal safety frameworks \citep{openai_preparedness_framework_2025}; \citep{anthropic_responsible_scaling_policy_2025}. Advanced models may also pose harm prior to public deployment, through risks of internal misuse, theft by threat actors, or harmful actions by autonomous systems \cite{stixAIClosedDoors2025a}; \citep{acharya_delaney_internal_ai_risks_2025}.

Catastrophic frontier AI risks are commonly grouped into three core technical domains: CBRN threats, advanced cyber-capability risks, and advanced autonomous-behavior risks \citep{FrontierMitigations2025b}. Beyond these, systemic contributors such as AI race dynamics and organizational risk significantly shape both the likelihood and severity of harm \citep{hendrycks_overview_catastrophic_ai_risks_2023}. Competitive race dynamics can accelerate deployment and reduce safety investment, while organizational failures, including weak access controls, inadequate oversight, and flawed safety processes, can materially increase the probability and impact of catastrophic outcomes. These organizational risks are particularly salient because they sit largely outside the scope of current model-level evaluations.

Assurance can play a central role in managing these risks. Assurance is defined as the provision of evidence-backed statements that an organization’s governance, risk management, and control processes are effective \citep{institute_of_internal_auditors_global_internal_audit_standard_2024}. Such assurance provides confidence to both internal stakeholders, such as boards of directors, and external stakeholders, including the public and regulators, that systemic risks are being adequately identified and controlled. Internal audit is the established organizational mechanism for delivering this form of enterprise-wide assurance. Internal audit provides independent evaluation of an organization’s governance, risk-management, and control processes and reports directly to the management body or board, supporting both independence and accountability \citep{institute_of_internal_auditors_global_internal_audit_standard_2024}. As a discipline, internal audit dates back to the 1930s and has developed formalized methods for risk-based assurance over successive decades \citealp{pickett_internal_auditing_handbook_2010}. It is now standard governance infrastructure across publicly listed firms, financial-services organizations, and public-sector institutions \citep{chartered_institute_of_internal_auditors_what_is_internal_audit_charter_2022}.

\citet{schuett_frontier_ai_developers_need_an_2024} was among the first to argue that internal audit could provide structured assurance over organizational AI risks that extend beyond model-level safety evaluations. Regulatory developments now reinforce this direction. The EU General-Purpose AI Code of Practice, to which many frontier AI developers have committed \citep{european_commission_code_of_practice_final_version_2025}, does not itself impose direct obligations on an external assurance provider, but instead requires the signatory to define, allocate, and document responsibility for the assessment of its systemic risk identification, prioritization, mitigation, monitoring, and control processes. The Code further requires that the results of these assessments be reported to the management body or an equivalent governance body, such as the board, which bears responsibility for oversight of systemic risk \citep{european_commission_code_of_practice_final_version_2025}. This formalizes expectations for organizationally embedded, board-level assurance over systemic AI risk, while leaving discretion over the sourcing and structure of the assurance function.

At present, however, frontier AI assurance remains largely confined to narrow technical domains. Developers commission external evaluations of specific model capabilities and safeguards, such as assessments by METR (Model Evaluation and Threat Research), an independent AI safety research organization that evaluates advanced AI systems for dangerous capabilities and misuse risks \citep{metrDetailsMETRsEvaluation2025}, Apollo Research’s evaluations of scheming and deception \citep{meinke_frontier_models_capable_incontext_scheming_2025}, and Irregular’s assessments of cyber-offensive capabilities \citep{irregular_evaluating_gpt_5_2025}. Additional assurance is obtained through security certifications such as ISO and SOC 2, typically disclosed via Trust Centers \citep{OpenAITrustPortal}; \citep{anthropic_trust_portal_2025}. While these mechanisms strengthen confidence in discrete technical claims, there is little evidence that assurance currently extends to enterprise-wide governance, organizational controls, or compliance with frontier safety frameworks \citep{homewoodThirdpartyComplianceReviews2025}. In particular, risks arising from pre-deployment internal use of frontier-level models remain largely outside the scope of current external assurance regimes \citealp{stixAIClosedDoors2025a}; \citep{acharya_delaney_internal_ai_risks_2025}.

Extending internal-audit-style assurance to frontier AI developers, however, presents both benefits and limitations. Frontier AI firms operate in regulatory environments where standards and evaluation methods remain nascent and regulatory efforts fragmented and reactive \citep{anderljungFrontierAIRegulation2024}. They also exhibit rapid deployment cycles, with training compute growing 4--5$\times$ per year and major model releases occurring roughly every 110 days \citep{epochaiDataAIBenchmarking2024}. Competitive pressures further incentivize speed: U.S.--China technological rivalry frames frontier AI leadership as a matter of national security and economic competitiveness \citep{williamsWinningDefiningContest2025}, and evidence suggests that both governments anticipate managing risks downstream rather than constraining development \emph{ex ante} \citep{zvenyhorodskyiAIActionPlans2025}. These dynamics heighten both the operational friction and the information-sensitivity risks associated with internal audit. These conditions directly shape the feasibility, timing, and design of internal audit for frontier AI developers. \citet{schuett_frontier_ai_developers_need_an_2024}, in his analysis of how internal audit could strengthen risk governance in frontier AI organizations, identifies potential tensions, including friction with deployment cycles, the need for highly specialized technical expertise, and the challenge of preserving genuine independence within fast-moving, founder-led organizations.

In this paper, we examine the practical design space for an internal audit function in frontier AI organizations, including the risk domains such a function could review; how it could be sourced; the frequency of review cycles; and the options for accessing and handling necessary information. For each dimension, we assess the associated benefits and limitations to support frontier AI developers in making informed decisions about whether, and how, to establish internal audit capability.

Specifically, we ask the following questions:
\begin{itemize}[leftmargin=*]
\item What risks should an internal audit function provide assurance about?
\item How should the internal audit function be sourced?
\item How frequently should internal audits be conducted?
\item What information should internal auditors access?
\end{itemize}
For each question, we evaluate a set of plausible options and assess their associated benefits and limitations in order to support frontier AI developers in making informed decisions about whether, and how, to establish internal audit capability.

\section{What risks should an internal audit function provide assurance about?}

Determining which risks an internal audit function should provide assurance over is a foundational design choice, as it directly shapes the relevance, credibility, and operational impact of internal audit within a frontier AI developer. Under a risk-based internal auditing model outlined in appendix \ref{appendix:riskbasedaudit}, assurance priorities are derived from the organization’s risk profile rather than from fixed, rotational coverage of business units \citep{pickett_internal_auditing_handbook_2010}. This process produces a board-approved audit plan proposed by the Chief Audit Executive (CAE), which specifies the functions, processes, systems, and governance arrangements to be reviewed over the coming months, and thereby defines the set of organizational risks over which internal audit will provide formal assurance.

In practice, each engagement on the audit plan is conducted using a defined type of internal audit review, depending on whether it examines controls relating to people, digital infrastructure, cybersecurity, regulatory compliance, or adherence to internal policy \citep{institute_of_internal_auditors_global_internal_audit_standard_2024}; \citep{pickett_internal_auditing_handbook_2010}; \citep{emporia_types_internal_audits_2025}. Applied to frontier AI developers, these conventional audit types can be re-aggregated into three safety-critical layers of assurance: model-level, system-level, and governance-level review (outlined in Table 1). These layers align traditional audit methodologies with the points at which catastrophic risk materializes in frontier AI systems: within the behavior of the model itself, the technical and organizational systems that surround it, and the governance processes that authorize, monitor, and constrain its use.

\paragraph{Model-level audits}

Model-level audits provide assurance that a frontier model’s observable behavior and latent capabilities align with defined safety objectives, including whether dangerous-capability evaluations reflect real behavior, whether classifiers reliably block harmful outputs under adversarial conditions, and whether training and alignment processes have meaningfully constrained dangerous capabilities \citep{anthropic_anthropic_s_responsible_scalin_2023,openai_gpt_4_system_card_2023}. Their primary advantage is directness: they assess the source of risk itself rather than relying on downstream compensating controls, making them a high-leverage intervention point whose effects propagate across deployments. While inability-based safety arguments have justified a strong focus on this layer to date \citep{clymer_safety_cases_2024}, crossing higher capability thresholds \citep{OpenAITrustPortal}; \citep{anthropic_claude_4_system_card_2025} exposes its limits. Classifiers and refusals exhibit measurable false-negative rates and edge cases \citep{wei_jailbroken_how_does_llm_safety_2024}, and assurance degrades as adversarial techniques evolve \citep{perez_red_teaming_language_models_wi_2022}, implying potential material residual risk at scale. Model-level controls also offer no protection against model-weight exfiltration \citep{nevoSecuringAIModel2024a}. These audits may be infeasible for rapidly iterated pre-deployment models \citep{stixAIClosedDoors2025a}; \citep{acharya_delaney_internal_ai_risks_2025} and can be high-friction for deployments, requiring scarce expertise and tight integration with training pipelines \citep{metr_evaluating_frontier_ai_r_d_cap_2024}. Nonetheless, many frontier developers already integrate third-party model evaluations into deployment pipelines, meaning that much of the required infrastructure and process for high-friction model reviews is already in place, reducing marginal cost and disruption. Overall, model-level audits offer the strongest direct signal on dangerous capabilities but entail the highest cost, fastest assurance decay, and greatest deployment friction.

\begingroup
\centering
\captionof{table}{Comparison of internal audit levels for AI safety assurance}
\label{tab:audit-levels}
\small
\setlength{\tabcolsep}{6pt}
\renewcommand{\arraystretch}{1.2}
\begin{tabular}{@{}p{0.10\textwidth} p{0.20\textwidth} p{0.19\textwidth} p{0.19\textwidth} p{0.19\textwidth}@{}}
\toprule
\textbf{Audit level} & \textbf{Description} & \textbf{Assurance degradation rate} & \textbf{Organizational cost} & \textbf{Deployment friction} \\
\midrule
\textbf{Model-level} & 
Evaluates a model’s behaviors and latent capabilities against safety objectives, including dangerous-capability tests and classifier reliability. & 
\cellcolor{HighRed}\textbf{High}
Degrades rapidly as jailbreaks, prompting strategies, and model architectures evolve, requiring frequent re-evaluation. & 
\cellcolor{HighRed}\textbf{High}
Requires scarce AI safety specialists, dedicated evaluation infrastructure, secure access to training and inference artifacts, and repeated testing. & 
\cellcolor{HighRed}\textbf{High}
Directly gates releases, may require evaluation re-runs, and interacts heavily with the ML development pipeline. \\
\addlinespace[1em]

\textbf{System-level} & 
Assesses whether surrounding technical and organizational systems can reliably detect and contain risks in real operational environments. & 
\cellcolor{MedYellow}\textbf{Medium}
More stable; core logging, access control, and monitoring controls evolve more slowly than models. & 
\cellcolor{MedYellow}\textbf{Medium}
Traditional IT, security, and resilience audit skills transfer effectively, with lower reliance on frontier ML expertise. & 
\cellcolor{MedYellow}\textbf{Medium}
Touches critical infrastructure but typically sits outside the primary model deployment critical path. \\
\addlinespace[1em]

\textbf{Governance-level} & 
Evaluates decision-rights, oversight mechanisms, escalation pathways, and responsible scaling governance. & 
\cellcolor{LowGreen}\textbf{Low}
Most stable; governance structures and escalation authorities change infrequently. & 
\cellcolor{LowGreen}\textbf{Low}
Established governance audit methodologies with minimal specialized technical tooling. & 
\cellcolor{LowGreen}\textbf{Low}
Adds procedural and reporting requirements but does not directly constrain technical systems. \\
\addlinespace[1em]

\textbf{Multi-level} & 
Integrates model, system, and governance audits to provide end-to-end assurance and reduce disconnects between true technical risk and organizational oversight. & 
\cellcolor{LowGreen}\textbf{Low}
Most resilient; degradation in any single layer can be detected or compensated by controls in other layers. & 
\cellcolor{HighRed}\textbf{High}
Highest total cost due to multidisciplinary scope, but with potential efficiency gains from integrated planning and shared evidence bases. & 
\cellcolor{MedYellow}\textbf{Medium}
Can be strategically sequenced across layers to minimize disruption to critical deployment paths. \\
\bottomrule
\end{tabular}
\endgroup

\paragraph{System-level audits}

System-level audits provide assurance that the technical and organizational systems surrounding frontier models can reliably detect, enforce, contain, and respond to emerging risks in practice, including model-weight security, access controls to unreleased models, usage monitoring, anomaly detection, API abuse prevention, post-deployment monitoring, and incident response \citep{nevoSecuringAIModel2024a}. Their main advantage is operational realism: they test whether controls function in live production environments rather than controlled evaluation settings, revealing whether logging captures relevant interactions, whether monitoring generates actionable alerts, and whether response teams can execute containment under real conditions. System-level audits also benefit from strong methodological transfer from mature IT and cybersecurity audit practices, including COBIT, the NIST Cybersecurity Framework, and IIA technology audit guides \citep{national_institute_of_standards_framework_for_improving_critic_2018}; \citep{isaca_cobit_2019_framework}; \citep{iia_global_technology_audit_guides_2024}, reducing execution cost and specialist dependency relative to model-level reviews. Their findings are also more temporally stable, as infrastructure and response capabilities evolve through deliberate system changes rather than continuous adversarial pressure. However, system-level audits alone cannot quantify the underlying level of model risk they are designed to contain: they assess whether jailbreaks would be detected and mitigated, but not how frequently they succeed. They also cannot validate whether deployment decisions were appropriate in light of the model’s true capabilities. Overall, system-level audits provide durable, operationally grounded assurance over enforcement and containment, but depend on model- and governance-level assurance to determine what level of risk those controls must absorb and whether deployment should proceed at all.

\paragraph{Governance-level audits.}
Governance-level audits provide assurance that human oversight, escalation mechanisms, and decision-rights for frontier AI systems operate independently and effectively, including board oversight of dangerous capabilities, deployment decision governance, compliance with responsible scaling commitments, and the integrity of external disclosures. The IIA defines governance evaluation as encompassing strategic decision-making, oversight of risk and control, promotion of ethics and values, and organizational performance management \citep{institute_of_internal_auditors_global_internal_audit_standard_2024}. The goal is to verify that the right decisions are made by the right people with the right information, independent of specific technical implementations. Governance-level auditing is also the lowest-cost and lowest-friction layer, relying primarily on interviews, document review, and process walkthroughs, and its findings are the most temporally stable because decision authorities and oversight structures change infrequently. However, governance assurance can become disconnected from technical reality. An organization may display formally sound oversight while underlying technical controls fail in practice, as illustrated by the 2017 Equifax breach, where investigations found that despite documented risk registers and incident-handling policies, known vulnerabilities were not patched and were subsequently exploited \citep{house_oversight_equifax_report_2018}; \citep{fca_equifax_final_notice_2023}. Governance audits alone therefore cannot validate the effectiveness of classifiers, weight security, or monitoring systems, and well-designed oversight processes may still fail under commercial or operational pressure without corroboration from system-level evidence. Overall, governance-level audits provide essential assurance over accountability and decision-rights, but are insufficient on their own to assure technical safety or operational containment.

\section{How should the internal audit function be sourced?}

Who performs internal audit work is a strategic choice for frontier AI labs: sourcing arrangements affect, among others, the credibility of assurances, access to scarce technical expertise, and protection of sensitive system assets. In practice, internal audit resourcing spans a spectrum: some organizations rely on fully in-house teams, others obtain internal audit services entirely from external providers, and many use combinations in between \citep{the_institute_of_internal_auditors_staffing_considerations_for_in_2018}; \citep{iia_global_view_internal_audit_2022}.

For frontier AI developers, the central trade-offs across sourcing models concern how to obtain the greatest assurance value, through access to specialized technical expertise and internal audit knowledge, coupled with meaningful independence, while managing the risks associated with granting access to sensitive system assets. The key trade-offs of sourcing models are outlined in Table 2.

\paragraph{In-house}

In-house internal audit functions are staffed primarily or entirely with employees of the organization, with the Chief Audit Executive (CAE) responsible for the audit charter, planning, execution, and reporting \citep{the_institute_of_internal_auditors_staffing_considerations_for_in_2018}. Independence is maintained through direct reporting lines to the board or audit committee, organizational separation from product and safety teams, and insulation from incentives tied to model development or deployment \citep{institute_of_internal_auditors_global_internal_audit_standard_2024}; \citep{bcbs_governance_framework_d383_2016}. The principal benefits of this model are continuity, institutional memory, and deep organizational understanding. Continuous proximity to systems and teams allows auditors to develop rich insight into how safety processes operate in practice and to track remediation over time \citep{homewoodThirdpartyComplianceReviews2025}. From an information-security perspective, in-house teams operate within the organization’s hardened infrastructure and clearance regimes, enabling controlled access to highly sensitive assets without external data transfer. However, fully internal teams may face challenges of perceived legitimacy, as external stakeholders can view their work as ``marking one’s own homework,'' particularly where findings lack independent visibility. This concern underpins recent proposals for structured third-party compliance reviews as complements to internal governance \citep{homewoodThirdpartyComplianceReviews2025}, and is illustrated by Anthropic’s decision to pair its internal sabotage-risk analysis with an external METR review \citep{metrDetailsMETRsEvaluation2025}; \citep{anthropic_sabotage_risk_report_internal_stress_testing_2025}. While internal teams can develop dual technical–audit expertise in principle, drawing auditors from technical or safety teams may compromise independence if staff audit controls they recently helped design, unless clear cooling-off safeguards are applied \cite{bcbs_governance_framework_d383_2016}. The in-house model therefore maximizes continuity and secure internal access, but trades these advantages against constraints on perceived external legitimacy and reliance on scarce hybrid talent.

\paragraph{Co-sourcing}

Co-sourcing retains an internal audit function that owns the charter, risk-based planning, reporting lines, and overall methodology, while specific audits or components are conducted jointly with external specialists \citep{the_institute_of_internal_auditors_staffing_considerations_for_in_2018}. Its central advantage is that it combines internal organizational context with external technical and methodological depth. The internal team anchors risk prioritization, process knowledge, and remediation tracking, while external experts contribute specialist capabilities in areas such as model evaluation, cybersecurity, or incident response that the internal function may lack \citep{homewoodThirdpartyComplianceReviews2025}. Co-sourcing also strengthens perceived independence: external participation in high-salience reviews signals impartiality, while preserving board-mandated authority within the internal audit function. From an information-security standpoint, it offers significant flexibility. Access to the most sensitive artifacts can remain restricted to internal auditors or even solely to the CAE, with external specialists relying on validated responses, structured checklists, or assumption-based assurance. This model is illustrated by METR’s external review of Anthropic’s Sabotage Risk Report, where assurance was based primarily on responses to detailed assumptions rather than direct inspection of underlying sensitive evidence \citep{metrDetailsMETRsEvaluation2025}. However, co-sourcing introduces coordination and governance complexity. Clear role delineation is required to avoid duplication, fragmented accountability, or ambiguity over ownership of findings and remediation. Secure information-sharing architectures must also be carefully engineered to manage disclosure risk. Overall, co-sourcing offers the strongest balance between independence, technical depth, and information security, at the cost of greater governance and coordination complexity.

\begingroup
\centering
\captionof{table}{Sourcing models and key trade-offs}
\label{tab:sourcing-models}
\small
\setlength{\tabcolsep}{6pt}
\renewcommand{\arraystretch}{1.2}

\begin{tabular}{@{}p{0.11\textwidth} p{0.19\textwidth} p{0.19\textwidth} p{0.19\textwidth} p{0.19\textwidth}@{}}
\toprule
\textbf{Sourcing model} &
\textbf{Risk to independence and external credibility} &
\textbf{Risk of insufficient technical and process expertise} &
\textbf{Risk to information security and data protection} &
\textbf{Cost, rigidity, and scalability risk} \\
\midrule

\textbf{In-house} &
\cellcolor{MedYellow}\textbf{Medium}
Strong structural independence is achievable, but purely internal assurance may lack external legitimacy unless complemented by visible third-party reviews. &
\cellcolor{MedYellow}\textbf{Medium}
Frontier AI developers have deep technical expertise in-house and the ability to attract top talent. However, they may need to build internal knowledge around running internal audits. &
\cellcolor{LowGreen}\textbf{Low}
Audit staff operate within existing high-security infrastructure and clearance regimes, minimizing cross-organizational data-transfer risks. &
\cellcolor{MedYellow}\textbf{Medium}
Higher fixed costs and longer time required to establish an internal audit team internally, but enables strong control over priorities and rapid re-tasking. \\
\addlinespace[1em]

\textbf{Co-sourced} &
\cellcolor{LowGreen}\textbf{Low}
Properly governed, co-sourced combines board-mandated organizational independence with visible participation of independent experts in high-stakes audits, boosting credibility. &
\cellcolor{LowGreen}\textbf{Low}
Enables an internal team to benefit from knowledge of internal systems and processes, while also drawing on diverse external specialists for tasks such as sabotage evaluations. &
\cellcolor{MedYellow}\textbf{Medium}
Requires carefully designed secure interfaces and contracts; access to highly sensitive data can be limited to internal auditors. &
\cellcolor{LowGreen}\textbf{Low}
Blends a fixed base with scalable external capacity, allowing intensive, episodic deep-dives without permanently expanding headcount. \\
\addlinespace[1em]

\textbf{Outsourced} &
\cellcolor{LowGreen}\textbf{Low}
External status can signal high independence, but commercial conflicts and ``wild west'' perceptions of AI audit markets can undermine trust if not tightly governed. &
\cellcolor{MedYellow}\textbf{Medium}
Vendors may have less technical expertise than internal teams and lack detailed knowledge of internal processes and technology. However, they have deep experience planning, running and reporting internal audits, and can provide insights from trends observed across other organizations. &
\cellcolor{HighRed}\textbf{High}
Sensitive data sharing is required outside of the organization or audits are likely to be constrained, limiting their value. &
\cellcolor{MedYellow}\textbf{Medium}
Faster setup time but may be expensive and additional overheads exist for repeated tendering and oversight. \\
\bottomrule
\end{tabular}
\endgroup

\paragraph{Outsourcing}

Under a fully outsourced model, an external provider supplies most or all internal audit personnel, with the organization retaining only a liaison or coordination role \citep{the_institute_of_internal_auditors_staffing_considerations_for_in_2018}. The primary advantages are rapid establishment, mature audit methodology, strong external benchmarking, and high perceived independence. External providers bring established documentation standards, board-reporting discipline, and cross-sector experience that may exceed the organizational maturity of early-stage frontier AI developers, particularly for system-level and governance-level audits using frameworks such as NIST, COBIT, ISO/IEC 27001, and the IIA Global Standards \citep{national_institute_of_standards_framework_for_improving_critic_2018,institute_of_internal_auditors_global_internal_audit_standard_2024}; \citep{isaca_cobit_2019_framework}. External validation can be seen as more independent than internal teams, with surveys showing greater stakeholder trust when assurance is external (\citep{}BSI, 2025; UK DSIT, 2024). Fully outsourced models can also be set up quickly, and then changed in line with the requirements of an organization, as illustrated by Monzo’s progression from fully outsourced to co-sourced and finally largely in-house audit functions between 2018 and 2024 \citep{monzo_annual_report_2018}; \citep{monzo_annual_report_2021}; \citep{monzo_annual_report_2022}; \citep{monzo_annual_report_2024}. However, outsourcing introduces heightened conflict-of-interest risks where providers also sell consultancy or tooling, a concern well documented in financial regulation \citep{bcbs_governance_framework_d383_2016}. External teams also lack deep knowledge of proprietary architectures and safety cultures, limiting their ability to assess whether controls operate in practice rather than merely on paper. From a security perspective, outsourcing requires the broadest external sharing of sensitive information, which may be impracticable for frontier labs with extreme misuse risk \citep{nevoSecuringAIModel2024a}. Outsourcing therefore maximizes perceived independence and speed of deployment, but does so at the expense of information-security exposure and deep operational insight.

Across all three models, sourcing choices reflect a fundamental trade-off between independence, technical depth, information security, and organizational control. In-house models maximize access and continuity but face credibility and talent constraints; outsourcing maximizes perceived independence and speed but raises security and contextual-knowledge risks; and co-sourcing offers a balanced middle ground while introducing coordination complexity. For frontier AI developers, retaining an internal Chief Audit Executive with the ability to orchestrate external specialists provides the greatest flexibility as assurance needs evolve with capability growth.

\section{How frequently should internal audits be conducted?}

The frequency of internal audit activity determines how current and relevant assurance remains. Assurance findings are time-bound: they reflect the state of controls at a particular point in time, and their relevance is likely to diminish as the organization's risk environment evolves. Two distinct frequency cycles shape this: how often the audit plan is updated, which determines the areas selected for upcoming review (planning frequency''), and how often individual areas within the plan are audited, which determines how frequently assurance is refreshed for those areas (execution frequency'').

The audit plan is the output of risk-based planning (see appendix \ref{appendix:riskbasedaudit}). It is a board-approved schedule of planned audit engagements, typically covering the next 12 months, that specifies which areas, processes, or systems have been prioritized for review, at what frequency, and why. Standards from the Institute of Internal Auditors (IIA), a global professional association for internal auditors, require that the risk assessment underlying the audit plan be performed at least annually, while accompanying guidance recommends reviewing it at least quarterly and updating it as needed in dynamic environments \citep{institute_of_internal_auditors_global_internal_audit_standard_2024}. If the materiality of risks changes but the plan is not updated, internal audit may be directing resources toward areas that no longer represent the highest priority. For example, over the span of three months, Anthropic's cyber threat landscape shifted materially: in August 2025, the company reported that cyberattacks using their model involved humans `very much in the loop directing operations'; by November 2025, it disclosed a state-sponsored campaign where ``the threat actor was able to use AI to perform 80--90

Within the plan, individual areas can be reviewed annually, quarterly, or continuously. The IIA does not prescribe fixed intervals, acknowledging that appropriate frequencies depend on organizational context, risk exposure, and resource constraints \citep{institute_of_internal_auditors_global_internal_audit_standard_2024}. Organizations may audit different areas at different frequencies based on assessed risk: some areas may be reviewed only every two to three years, while others are audited annually or more frequently. Execution frequency that is too low allows control weaknesses to persist undetected, leaving the organization exposed to risks it believes are controlled; frequency that is too high drives escalating costs and ``assurance fatigue'' \citep{schuett_frontier_ai_developers_need_an_2024}. For frontier AI developers, the key question is how often high-risk areas should be audited given the pace of change in their operating environment.

This section discusses options for execution frequency (outlined in table 3), evaluating different positions on the spectrum between assurance value and organizational cost.

\paragraph{Annual cycles}

Annual audits represent the standard execution frequency across industries: high-risk areas are typically reviewed annually, moderate-risk areas every 12--24 months, and low-risk areas every 24--36 months \citep{institute_of_internal_auditors_global_internal_audit_standard_2024}. For frontier AI developers, annual cycles provide a baseline level of assurance and a sustainable balance between coverage and operational friction. One benefit is predictability: annual reviews create a stable oversight cadence and maintain governance expectations without overwhelming teams. However, frontier AI environments evolve far faster than traditional sectors. Training compute, a proxy for model capability, appears to double roughly every six months \citep{giattino_samborska_training_compute_2025}, and major organizational restructuring, such as Anthropic doubling its workforce to 1,300 by 2025 \citep{robison_anthropic_first_developer_conference_2025}, can alter risk profiles long before the next scheduled audit, especially if it involves the creation of new teams and departments, causing significant deviations from established processes and reporting structures. As a result, audit findings may lose relevance months before the next cycle, reducing assurance value and allowing deviations in safety processes to persist undetected. Annual cycles are therefore best suited to governance-level audits and stable system-level controls, where oversight structures and core infrastructure change relatively slowly compared to model capabilities. The overall trade-off is that annual audits are sustainable and professionally standard but may not maintain assurance relevance in the face of changing model capabilities and organizational dynamics.

\begingroup
\centering
\captionof{table}{Internal audit execution frequency options}
\label{tab:execution-frequency}
\small
\setlength{\tabcolsep}{6pt}
\renewcommand{\arraystretch}{1.2}

\begin{tabular}{@{}>{\raggedright\arraybackslash\sloppy}p{0.15\textwidth}
                >{\raggedright\arraybackslash\sloppy}p{0.25\textwidth}
                >{\raggedright\arraybackslash\sloppy}p{0.26\textwidth}
                >{\raggedright\arraybackslash\sloppy}p{0.25\textwidth}@{}}
\toprule
\textbf{Approach} & \textbf{Advantages} & \textbf{Disadvantages} & \textbf{Recommendation} \\
\midrule

\textbf{Annual cycles} &
Provides a professional standard across industries and a sustainable balance between assurance and operational efficiency. &
Findings risk becoming outdated between cycles given rapid capability advances and organizational growth in frontier AI. &
Represents the baseline professional frequency for internal audit. \\
\addlinespace[1em]

\textbf{Semi-annual or quarterly cycles} &
Provides more current assurance and reduces the risk that significant deviations persist undetected. &
Substantial resource requirements; uncommon across industries; may be difficult to sustain. &
Use following significant incidents or for highest-risk areas in frontier AI organizations seeking more ambitious assurance. \\
\addlinespace[1em]

\textbf{Ad hoc reviews (rapid assurance)} &
Enables fast response to emerging risks not in the scheduled audit plan; provides timely limited assurance within 6--8 weeks. &
Provides limited rather than comprehensive assurance; requires established rapid-assurance methodologies and capabilities. &
Recommended for organizations seeking more ambitious assurance and requiring flexible rapid-response audit capacity. \\
\addlinespace[1em]

\textbf{Continuous auditing} &
Delivers real-time or near real-time issue identification, enabling prompt response and supporting regulatory compliance (e.g., EU AI Act). &
Requires substantial upfront investment, specialist technical expertise, robust data-security safeguards, and oversight of automated audit systems. &
A valuable complement to periodic audits; comprehensive programs may invest heavily in continuous auditing as the pace of change increases. \\
\bottomrule

\end{tabular}
\endgroup
\paragraph{Semi-annual or quarterly cycles}

Semi-annual or quarterly cycles involve re-auditing selected high-risk areas more than once per year, providing more current assurance and narrowing the window in which control failures or capability shifts can persist unnoticed. Such cadences are rarely documented even in high-risk industries, largely because formal internal audits require substantial preparation, fieldwork, and reporting, often spanning several weeks and demanding significant time from both auditors and auditees \citep{institute_of_internal_auditors_global_internal_audit_standard_2024}. In frontier AI, the limiting factor is not only internal-audit capacity but the cumulative burden on operational teams: more frequent audits require additional interviews, documentation requests, walkthroughs, and remediation tracking, diverting attention from core safety, engineering, and governance work. Nevertheless, semi-annual or quarterly cycles may be appropriate when material events indicate the need for repeated follow-up, such as major control failures, regulatory enforcement actions, or data breaches in other sectors \citep{martin_integrated_approach_security_audits_2022}. For frontier AI developers, analogous triggers include safety incidents, unexpected capability gains, or the crossing of pre-defined evaluation thresholds indicating heightened risk. In these contexts, a scheduled increase in audit cadence can help confirm that remediation remains effective across multiple cycles before further model changes or deployments proceed. This frequency is therefore most suitable for model-level audits and the fastest-moving system-level controls, where assurance can become outdated quickly due to capability advancement or evolving threat models. The overall trade-off is that although more frequent audits improve temporal sensitivity, they substantially increase friction for audited teams and must therefore be applied selectively to areas where rapid change or high stakes justify the added organizational burden.

\paragraph{Ad hoc reviews}

Ad hoc reviews are unscheduled assurance engagements initiated in response to emerging risks, incidents, or material changes outside the normal audit plan. They are appropriate when an area encounters an unexpected shock, such as a safety incident, model-behavior anomaly, or governance breakdown, or when rapid organizational change introduces risks not anticipated during planning. They can apply to any audit layer depending on the trigger, for example, a newly discovered jailbreak class or refusal-drift for model-level audits, monitoring or logging anomalies for system-level audits, or deployment decisions taken without required safety sign-offs for governance-level audits. Unlike semi-annual or quarterly cycles, which provide repeated follow-up for areas already on the plan, ad hoc reviews offer one-off, immediate assurance when timely scrutiny is required but the affected area is not scheduled for near-term review. Established methodologies such as rapid assurance reviews support this approach, providing focused, time-bounded assurance over emerging processes or technologies within 6--8 weeks while adhering to professional standards \citep{global_fund_rapid_assurance_review_gc7_2025}. The key advantage is responsiveness: they allow frontier AI developers to direct assurance precisely where it is needed, at the moment it is needed, rather than waiting for the next cycle. The drawbacks are opportunity costs and operational disruption, as unplanned reviews can divert audit resources and impose short-notice demands on teams already managing incidents or implementing remediation. The trade-off is that ad hoc reviews provide agility but can strain both audit and operational teams if overused; they are therefore best reserved for genuinely material developments that warrant immediate assurance.

\paragraph{Continuous audit}

Continuous audit refers to methodologies in which control and risk assessments are performed automatically on an ongoing basis \citep{coderre_continuous_auditing_gtag_2006}, rather than at scheduled intervals. It relies heavily on automated data collection and analysis \citep{iia_gtag_continuous_auditing_monitoring_2025} and, when implemented effectively, improves the reliability of information available to decision makers while detecting emerging risks and control weaknesses in near real time \citep{iia_gtag_continuous_auditing_monitoring_2025}. For frontier AI developers, continuous auditing offers two major benefits: timeliness, as issues can be surfaced as they arise rather than months later; and reduced burden on auditees, since automated pipelines can supply evidence without requiring repeated manual requests for logs, reports, or walkthroughs. These features align with regulatory expectations for continuous post-market monitoring under the EU AI Act \citep{european_commission_code_of_practice_final_version_2025} and academic work noting that such methods can "keep pace with the AI system’s evolution" \citep{minkkinen_continuous_auditing_ai_2022}. However, continuous auditing requires upfront investment in monitoring infrastructure, sustained engineering maintenance, and specialized digital skills often lacking in internal audit teams \citep{betti_sarens_internal_audit_digitalised_2021}. Large volumes of sensitive data must also be handled securely, increasing information-security risk. Moreover, automated systems may create a potential "infinite regress", as the audit tools themselves require oversight \citep{minkkinen_continuous_auditing_ai_2022}. Continuous auditing is most appropriate for system-level controls and certain model-adjacent technical signals where assurance degrades rapidly and real-time visibility is critical. The trade-off is that while continuous audit provides the highest temporal sensitivity with minimal disruption to teams, it is the most technically demanding and least mature option, requiring careful design and sustained investment.

Annual audits provide the professional baseline for high-risk areas, but in frontier AI the relevance of findings can degrade far more quickly than internal audit capacity can revisit them. Yet audit frequency is constrained not only by internal audit resources but by auditee bandwidth: repeatedly reviewing the same high-intensity areas can impose significant disruption on technical and safety teams unless carefully scoped and sequenced. Overall, the trade-off is between assurance freshness and organizational load: frontier AI developers will need to balance annual structural reviews, selective higher-frequency audits for the most dynamic risks, and continuous monitoring for system-level controls where automation can meaningfully reduce auditee disruption.

\section{What information should internal auditors access?}

Internal auditors are expected to have unrestricted access to all information necessary to conduct their assurance activities, including data, records, systems, property, and personnel, in order to provide accurate, evidence-based conclusions \citep{institute_of_internal_auditors_global_internal_audit_standard_2024}. However, in the context of frontier AI development, expanding access to highly sensitive assets, such as model weights or proprietary training pipelines, can introduce its own set of risks, including the potential for catastrophic misuse or national security threats \citep{nevoSecuringAIModel2024a}.

The level and type of information required is dependent on the type of audit being undertaken and the assurance being sought: providing reasonable assurance requires a minimum threshold of access, below which only limited assurance can be offered \citep{institute_of_internal_auditors_global_internal_audit_standard_2024}. \citet{homewood_third_party_compliance_reviews_2025} classify the information used to support third-party compliance reviews of frontier AI safety frameworks into four categories: structural, procedural, operational, and technical. We adopt these categories, aligned with established audit-evidence standards \citep{institute_of_internal_auditors_global_internal_audit_standard_2024}, to evaluate the minimum information access required for internal audit to provide reasonable assurance at each audit layer, distinguishing between information needed for risk-based audit planning and for audit fieldwork, as well as the trade-offs and practical options for enabling such access without creating a high information-security risk (outlined in Table 4).

\begingroup
\centering
\captionof{table}{Information types: applicability and key trade-offs}
\label{tab:information-types}
\small
\setlength{\tabcolsep}{6pt}
\renewcommand{\arraystretch}{1.2}

\begin{tabular}{@{}
>{\raggedright\arraybackslash\sloppy}p{0.23\textwidth}
>{\raggedright\arraybackslash\sloppy}p{0.30\textwidth}
>{\raggedright\arraybackslash\sloppy}p{0.40\textwidth}
@{}}
\toprule
\textbf{Information type} &
\textbf{Essential for} &
\textbf{Key trade-offs} \\
\midrule

\textbf{Structural} (org charts, governance artifacts, policies) &
Risk-based planning; governance-level audits. &
Low security risk, but restricting access undermines audit planning and governance assurance. Provides stated intent rather than evidence of real behavior. \\
\addlinespace[1em]

\textbf{Procedural} (procedures, system documentation, evaluation methods) &
Risk-based planning; system-level audits; supports model-level and governance-level audits. &
Moderate security risk; exposing control logic may aid adversaries. Restricted access limits assurance to policy intent rather than effectiveness. \\
\addlinespace[1em]

\textbf{Operational} (interviews, walkthroughs, incident records) &
Governance-level and system-level audits focused on operating effectiveness. &
Medium to high sensitivity; without operational access, audits cannot test real-world control effectiveness. \\
\addlinespace[1em]

\textbf{Technical} (model weights, logs, evaluation artifacts, infrastructure details) &
System-level and model-level audits. &
Highest security risk; restricting access removes model-level assurance and weakens system assurance, while full access increases exposure. \\
\bottomrule
\end{tabular}
\endgroup

\paragraph{Structural information}

Structural information includes organizational charts, board minutes, governance frameworks, and other artifacts describing how an organization intends to govern safety \citep{homewoodThirdpartyComplianceReviews2025}. It has relatively low information-security sensitivity, since much of it is public or near-public, though internal governance documents can carry reputational risk if disclosed. Structurally, this information is essential for risk-based audit planning (detailed in Appendix \ref{appendix:riskbasedaudit}): it enables auditors to map reporting lines, identify auditable units, and understand who holds formal accountability for catastrophic-risk decisions. It is also foundational for governance-level audits, forming the core evidence base for assessing board oversight, escalation pathways, and the distribution of decision-rights. By contrast, structural information provides limited direct evidence for model-level or system-level audits: it reveals governance intent rather than technical performance or operational reality. The trade-off is asymmetric: restricting access undermines accurate audit planning and renders governance audits superficial, while full access carries only modest reputational exposure, making structural information the clearest case for unrestricted auditor access.

\paragraph{Procedural information}

Procedural information includes internal procedures, system documentation, technical specifications, and logs describing how policies are operationalized \citep{homewoodThirdpartyComplianceReviews2025}. This category carries moderate security sensitivity, since disclosure of control logic or system architecture may facilitate targeted bypass by adversaries. Procedural information allows auditors to assess whether controls are designed adequately to operationalize stated policies and achieve risk management objectives. It is core evidence for system-level audits, which rely on it to test access controls, monitoring configurations, incident-response processes, and change-management pathways. It also supports model-level audits by revealing how evaluations and mitigations are intended to function, and it is important for governance audits because it demonstrates how high-level decisions translate into operational practice. In some cases, procedural information alone can support meaningful assurance without direct access to models. METR's methodology review of OpenAI's gpt-oss-120b pre-release evaluations illustrates this approach: METR reviewed the methodology OpenAI used to evaluate whether the model could be adversarially fine-tuned to reach dangerous capability thresholds, producing recommendations based on procedural documentation alone. This enabled a bounded form of assurance over evaluation methodology adequacy without requiring access to model weights or direct technical testing \citep{metrDetailsMETRsEvaluation2025}. The trade-off is that limiting access to procedural information forces auditors to assess controls only at the level of policy statements, preventing rigorous assessment of control design adequacy; but granting broad access can expose the defensive architecture of the organization. Frontier AI developers therefore must balance the benefits of verifiability against the heightened risk of enabling targeted adversarial bypass. Restricting procedural information confines auditors to assessing controls against high-level policies rather than internal standards and operating procedures, limiting assurance to control intent rather than design adequacy. Full disclosure, however, risks expanding the attack surface: adversaries who understand specific safeguards can mount targeted bypass attempts.

\paragraph{Operational information}

Operational information provides insight into how controls work in practice and includes interviews, process walkthroughs, incident records, and operational communications \citep{homewoodThirdpartyComplianceReviews2025}. It is considered medium-to-high sensitivity \citep{homewoodThirdpartyComplianceReviews2025} as it often contains tacit organizational knowledge, implicit control weaknesses, escalation failures, and internal communications that would be highly damaging if exposed. Unlike structural or procedural documents, operational artifacts may reveal how teams actually adapt, bypass, or reinterpret controls under pressure, which can disclose sensitive failure modes or security-relevant behaviors. For the same reason, operational evidence is essential for verifying operating effectiveness \citep{institute_of_internal_auditors_global_internal_audit_standard_2024}: it is the only way to detect divergence between documented processes and real-world behavior across all types of audit. METR’s review of Anthropic’s pilot Sabotage Risk Report demonstrates an approach for using operational information for assurance: Anthropic provided materials through written responses, interviews, and responses to a structured assurance checklist, allowing METR to rate where evidence was strong or limited \citep{metrDetailsMETRsEvaluation2025}. The limitations of excluding operational information are clear in cases such as the Boeing 737 MAX audits: documented procedures existed, but operational audits revealed systematic failure to follow them, contributing to fatal incidents \citep{miller_mintz_boeing_corporate_culture_2025}; \citep{faa_updates_boeing_737_9_max_2024}FAA, 2024. The core trade-off is that without operational information, audits are limited to assessing control design rather than operating effectiveness: they can confirm that policies exist and are appropriately specified, but cannot verify whether controls function as intended in practice \citep{institute_of_internal_auditors_global_internal_audit_standard_2024}.

\paragraph{Technical information}

Technical information includes model weights, evaluation results, system logs, training procedures, infrastructure configurations, and other artifacts that directly expose model behavior and system security \citep{homewoodThirdpartyComplianceReviews2025}. It is the most sensitive information category: weight exfiltration or capability-evaluation leakage could enable catastrophic harm \citep{nevoSecuringAIModel2024a}. Technical information enables auditors to assess whether automated and technical controls operate effectively in practice, providing direct evidence of system behavior rather than relying on self-reported compliance. Access is essential for model-level and system-level audits, enabling verification of logging completeness, access-control integrity, and security configurations, but less critical for governance audits. The Apollo Research collaboration with OpenAI illustrates how access to highly sensitive technical information can be structured safely. To evaluate scheming behavior, Apollo was granted rate-limited, time-bounded access to internal chain-of-thought traces for o3 and o4-mini, enabling inspection of behaviorally relevant signals while tightly constraining what could be accessed and under what monitoring conditions \citep{openai_detecting_reducing_scheming_2025}; \citep{schoen_stress_testing_deliberative_alignment_2025}. The trade-off is significant: restricting technical access eliminates model-level assurance and severely weakens system-level assurance, while full access poses the highest information-security risk. However, clean-room environments, tiered access, rate-limited engagements, and comprehensive session monitoring can materially reduce risk (detailed in Appendix \ref{appendix:accesscontrols}). Internal or co-sourced sourcing models further expand scope for proportionate access by keeping auditors within the organization's security perimeter. Where technical access cannot be provided even with such controls, internal audit should explicitly provide only limited assurance and articulate how evidential constraints affect its conclusions \citep{institute_of_internal_auditors_global_internal_audit_standard_2024}.

The strength of internal audit assurance is fundamentally determined by the information to which auditors have access. Restricting operational or technical information forces audits to assess control design rather than real-world effectiveness, and in the case of technical restrictions, may eliminate meaningful model- or system-level assurance altogether. Frontier AI developers must therefore balance information-security risks against assurance gaps, using proportionate safeguards, such as clean rooms, tiered access, and monitored, time-bounded sessions, to enable access without resorting to binary grant-or-deny decisions.

\section{Conclusion}

This paper has made the case for internal audit as a mechanism for allocating responsibility for systemic risk assessment in frontier AI companies, explored the practical trade-offs that shape its effectiveness, and provided insights to inform implementation of commitment 8 of the EU General-Purpose AI Code of Practice. The benefits of internal audit include the ability to scope assurance across model-level, system-level, and governance-level controls, and the ability to provide ongoing oversight as frontier AI systems evolve rapidly. Despite these benefits, internal audit faces real challenges: information-security risks if sensitive technical artefacts are shared beyond existing security perimeters, potential independence concerns depending on sourcing arrangements, and operational friction if audit frequencies impose excessive burden on development and safety teams. Nonetheless, we believe established audit and assurance practices adapted from other high-stakes industries can provide a basis for exploring solutions to these challenges for frontier AI developers.

Our analysis addresses four key dimensions of internal audit implementation for frontier AI: what scope of assurance internal audit can provide beyond model-level evaluations, how access to technical information affects assurance quality, what sourcing models can balance information security with independence requirements, and how audit frequencies can maintain relevance without creating excessive organizational friction. For each dimension, we describe trade-offs between different implementation approaches that reflect varying balances between assurance quality, security risk, independence considerations, and operational burden. Each organization can adapt its approach to match its governance maturity, risk appetite, and security posture. Over time, companies can build toward more comprehensive approaches by testing and implementing the mitigations for challenges identified in this analysis.

The memo leaves several questions unanswered that warrant further research and experimentation. In particular, more work is needed to understand optimal sourcing arrangements for frontier AI internal audit, as the unique combination of technical complexity, information sensitivity, and independence requirements may require novel hybrid models not yet established in other industries. Future work could examine how internal audit methodologies can be adapted to assess emerging control types specific to frontier AI, such as evaluations for model behaviours or governance mechanisms for model weight security. An examination of how continuous auditing and automated evidence collection can be practically implemented in frontier AI development environments would also be valuable. Finally, future research may help clarify the respective roles of internal audit, external evaluations, internal safety teams, and board oversight in creating a comprehensive assurance system for frontier AI. Of course, as frontier AI companies gain experience implementing internal audit functions, their practical insights will shape further research directions.

The AI governance field is not alone in its challenge to gain meaningful assurance over complex, high-stakes systems: established internal audit practices from financial services, healthcare, and critical infrastructure offer a strong foundation to build upon. Implementing internal audit functions can help frontier AI companies fulfill their responsibility for systemic risk assessment under the EU General-Purpose AI Code of Practice, helping strengthen risk management and board oversight capabilities. By adapting these practices to the unique characteristics of frontier AI systems, companies not only meet emerging regulatory expectations but also demonstrate a commitment to responsible development. Through proactive investment in internal audit capabilities, frontier AI developers can better prepare for future governance requirements and demonstrate leadership in frontier AI assurance.
\clearpage
\bibliographystyle{plainnat}
\bibliography{refs}

@misc{acharya_delaney_internal_ai_risks_2025,
  title        = {Managing Risks from Internal AI Systems},
  author       = {Acharya, Ashwin and Delaney, Oscar},
  year         = {2025},
  month        = jul,
  day          = {21},
  url          = {https://static1.squarespace.com/static/64edf8e7f2b10d716b5ba0e1/t/687e324254b8df665abc5664/1753100867033/Managing+Risks+from+Internal+AI+Systems.pdf},
  howpublished = {\url{https://static1.squarespace.com/static/64edf8e7f2b10d716b5ba0e1/t/687e324254b8df665abc5664/1753100867033/Managing+Risks+from+Internal+AI+Systems.pdf}},
  note         = {Institute for AI Policy and Strategy (IAPS)}
}

@article{robison_anthropic_first_developer_conference_2025,
  author       = {Robison, K.},
  title        = {Anthropic's First Developer Conference Was All About AI Agents},
  journaltitle = {WIRED},
  year         = {2025},
  month        = may,
  day          = {23},
  url          = {https://www.wired.com/story/anthropic-first-developer-conference/},
  howpublished = {\url{https://www.wired.com/story/anthropic-first-developer-conference/}},
  note         = {Accessed 30 November 2025}
}

@misc{martin_integrated_approach_security_audits_2022,
  author       = {Martin, C.},
  title        = {An Integrated Approach to Security Audits},
  year         = {2022},
  url          = {https://www.isaca.org/resources/news-and-trends/industry-news/2022/an-integrated-approach-to-security-audits},
  howpublished = {\url{https://www.isaca.org/resources/news-and-trends/industry-news/2022/an-integrated-approach-to-security-audits}},
  note         = {ISACA. Accessed 30 November 2025}
}

@techreport{global_fund_rapid_assurance_review_gc7_2025,
  author       = {{The Global Fund Office}},
  title        = {Rapid Assurance Review: Reduction of GC7 Country Allocations},
  year         = {2025},
  institution  = {{The Global Fund}},
  number       = {GF-OIG-25-013},
  address      = {Geneva},
  url          = {https://www.theglobalfund.org/media/u0blrt00/oig_gf-oig-25-013_report_en.pdf},
  howpublished = {\url{https://www.theglobalfund.org/media/u0blrt00/oig_gf-oig-25-013_report_en.pdf}},
  note         = {Accessed 30 November 2025}
}

@techreport{coderre_continuous_auditing_gtag_2006,
  author       = {Coderre, David},
  title        = {Global Technology Audit Guide: Continuous Auditing—Implications for Assurance, Monitoring, and Risk Assessment},
  year         = {2006},
  institution  = {{The Institute of Internal Auditors}},
  address      = {Montvale, NJ}
}

@misc{iia_gtag_continuous_auditing_monitoring_2025,
  author       = {{Institute of Internal Auditors}},
  title        = {Global Technology Audit Guide: Continuous Auditing and Monitoring},
  year         = 2025,
  url          = {https://www.theiia.org/en/content/guidance/recommended/supplemental/gtags/continuous-auditing-and-monitoring/},
  howpublished = {\url{https://www.theiia.org/en/content/guidance/recommended/supplemental/gtags/continuous-auditing-and-monitoring/}},
  note         = {IIA Global Technology Audit Guide (GTAG)}
}

@article{minkkinen_continuous_auditing_ai_2022,
  author       = {Minkkinen, Matti and Laine, Joakim and M{\"a}ntym{\"a}ki, Matti},
  title        = {Continuous Auditing of Artificial Intelligence: A Conceptualization and Assessment of Tools and Frameworks},
  journaltitle = {Digital Society},
  year         = {2022},
  volume       = {1},
  pages        = {21},
  doi          = {10.1007/s44206-022-00022-2},
  url          = {https://doi.org/10.1007/s44206-022-00022-2}
}

@article{betti_sarens_internal_audit_digitalised_2021,
  author       = {Betti, Nicola and Sarens, Gerrit},
  title        = {Understanding the Internal Audit Function in a Digitalised Business Environment},
  journaltitle = {Journal of Accounting \& Organizational Change},
  year         = {2021},
  volume       = {17},
  number       = {2},
  pages        = {197--216},
  doi          = {10.1108/JAOC-11-2019-0114},
  url          = {https://doi.org/10.1108/JAOC-11-2019-0114}
}

@article{miller_mintz_boeing_corporate_culture_2025,
  author       = {Miller, William F. and Mintz, Steven},
  title        = {The Story of Boeing’s Failed Corporate Culture},
  journaltitle = {The CPA Journal},
  year         = {2025},
  month        = jun,
  day          = {02},
  url          = {https://www.cpajournal.com/2025/06/02/the-story-of-boeings-failed-corporate-culture/},
  howpublished = {\url{https://www.cpajournal.com/2025/06/02/the-story-of-boeings-failed-corporate-culture/}},
  note         = {Accessed 30 November 2025}
}

@misc{faa_updates_boeing_737_9_max_2024,
  author       = {{Federal Aviation Administration}},
  title        = {Updates on Boeing 737-9 MAX Aircraft},
  year         = {2024},
  url          = {https://www.faa.gov/newsroom/updates-boeing-737-9-max-aircraft},
  howpublished = {\url{https://www.faa.gov/newsroom/updates-boeing-737-9-max-aircraft}},
  note         = {FAA Newsroom updates following January 2024 in-flight incident}
}

@misc{openai_detecting_reducing_scheming_2025,
  author       = {{OpenAI}},
  title        = {Detecting and Reducing Scheming in AI Models},
  year         = {2025},
  url          = {https://openai.com/index/detecting-and-reducing-scheming-in-ai-models/},
  howpublished = {\url{https://openai.com/index/detecting-and-reducing-scheming-in-ai-models/}}
}

@article{schoen_stress_testing_deliberative_alignment_2025,
  title        = {Stress Testing Deliberative Alignment for Anti-Scheming Training},
  author       = {Schoen, Bronson and Nitishinskaya, Evgenia and Balesni, Mikita and Højmark, Axel and Hofst{\"a}tter, Felix and Scheurer, J{\'e}r{\'e}my and Meinke, Alexander and Wolfe, Jason and van der Weij, Teun and Lloyd, Alex and Goldowsky-Dill, Nicholas and Fan, Angela and Matveiakin, Andrei and Shah, Rusheb and Williams, Marcus and Glaese, Amelia and Barak, Boaz and Zaremba, Wojciech and Hobbhahn, Marius},
  journal      = {arXiv preprint},
  volume       = {arXiv:2509.15541},
  year         = {2025},
  doi          = {10.48550/arXiv.2509.15541},
  url          = {https://arxiv.org/abs/2509.15541}
}

@article{anderljungFrontierAIRegulation2024,
  author       = {Anderljung, Markus and Korinek, Anton},
  title        = {Frontier {{AI Regulation}}: {{Safeguards Amid Rapid Progress}}},
  publisher    = {Lawfare},
  year         = {2024},
  url          = {https://www.lawfaremedia.org/article/frontier-ai-regulation-safeguards-amid-rapid-progress},
  journaltitle = {Lawfare},
  shorttitle   = {Frontier {{AI Regulation}}}
}

@misc{anthropic_claude_4_system_card_2025,
  title        = {Claude Opus 4 and Claude Sonnet 4 System Card},
  author       = {{Anthropic}},
  year         = {2025},
  month        = may,
  url          = {https://www-cdn.anthropic.com/6d8a8055020700718b0c49369f60816ba2a7c285.pdf},
  howpublished = {\url{https://www-cdn.anthropic.com/6d8a8055020700718b0c49369f60816ba2a7c285.pdf}}
}

@article{hendrycks_overview_catastrophic_ai_risks_2023,
  title        = {An Overview of Catastrophic AI Risks},
  author       = {Hendrycks, Dan and Mazeika, Mantas and Woodside, Thomas},
  journal      = {arXiv preprint},
  volume       = {arXiv:2306.12001},
  year         = {2023},
  url          = {https://arxiv.org/abs/2306.12001},
  doi          = {10.48550/arXiv.2306.12001}
}

@misc{epochaiDataAIBenchmarking2024,
  author       = {{Epoch AI}},
  title        = {Data on {{AI Benchmarking}}},
  year         = {2024},
  howpublished = {\url{https://epoch.ai/benchmarks}},
  note         = {Accessed 2025-11-25}
}

@misc{FrontierMitigations2025b,
  author       = {{Frontier Model Forum}},
  title        = {Frontier {{Mitigations}}},
  year         = {2025},
  howpublished = {\url{https://www.frontiermodelforum.org/technical-reports/frontier-mitigations/}},
  note         = {Accessed 2025-09-01}
}

@misc{homewoodThirdpartyComplianceReviews2025,
  author       = {Homewood, Aidan and Williams, Sophie and Dreksler, Noemi and Lidiard, John and Murray, Malcolm and Heim, Lennart and Ziosi, Marta and {hÉigeartaigh}, Seán Ó and Chen, Michael and Wei, Kevin and Winter, Christoph and Brundage, Miles and Garfinkel, Ben and Schuett, Jonas},
  title        = {Third-Party Compliance Reviews for Frontier {{AI}} Safety Frameworks},
  year         = {2025},
  howpublished = {\url{http://arxiv.org/abs/2505.01643}},
  note         = {Accessed 2025-08-16}
}

@misc{metrDetailsMETRsEvaluation2025,
  author       = {METR},
  title        = {Details about {{METR}}’s Evaluation of {{OpenAI GPT-5}}},
  year         = {2025},
  howpublished = {\url{https://evaluations.metr.org//gpt-5-report/}},
  note         = {Accessed 2025-11-25}
}

@misc{MITAIRisk,
  title        = {The {{MIT AI Risk Repository}}},
  year         = {2025},
  howpublished = {\url{https://airisk.mit.edu/}},
  note         = {Accessed 2025-12-12}
}

@report{nevoSecuringAIModel2024a,
  author       = {Nevo, Sella and Lahav, Dan and Karpur, Ajay and Bar-On, Yogev and Bradley, Henry Alexander and Alstott, Jeff},
  title        = {Securing {{AI Model Weights}}: {{Preventing Theft}} and {{Misuse}} of {{Frontier Models}}},
  year         = {2024},
  url          = {https://www.rand.org/pubs/research_reports/RRA2849-1.html},
  shorttitle   = {Securing {{AI Model Weights}}}
}

@misc{OpenAITrustPortal,
  author       = {OpenAI},
  title        = {{{OpenAI Trust Portal}} | {{Powered}} by {{SafeBase}}},
  year         = {2025},
  howpublished = {\url{https://trust.openai.com/}},
  note         = {Accessed 2025-12-04}
}

@misc{irregular_evaluating_gpt_5_2025,
  title        = {Evaluating GPT-5},
  author       = {{Irregular}},
  year         = {2025},
  url          = {https://www.irregular.com/publications/evaluating-gpt-5},
  howpublished = {\url{https://www.irregular.com/publications/evaluating-gpt-5}}
}

@book{pickett_internal_auditing_handbook_2010,
  author       = {Pickett, K. H. Spencer},
  title        = {The Internal Auditing Handbook},
  year         = {2010},
  publisher    = {John Wiley \& Sons},
  url          = {https://jabatanfungsionalauditor.wordpress.com/wp-content/uploads/2016/06/the-internal-auditing-handbook-k-h-spencer-pickett-2010.pdf},
  howpublished = {\url{https://jabatanfungsionalauditor.wordpress.com/wp-content/uploads/2016/06/the-internal-auditing-handbook-k-h-spencer-pickett-2010.pdf}}
}

@misc{openai_preparedness_framework_2025,
  title        = {OpenAI Preparedness Framework},
  author       = {{OpenAI}},
  year         = {2025},
  url          = {https://openai.com/safety/preparedness/},
  howpublished = {\url{https://openai.com/safety/preparedness/}}
}

@misc{anthropic_responsible_scaling_policy_2025,
  title        = {Responsible Scaling Policy},
  author       = {{Anthropic}},
  year         = {2025},
  url          = {https://www-cdn.anthropic.com/872c653b2d0501d6ab44cf87f43e1dc4853e4d37.pdf},
  howpublished = {\url{https://www-cdn.anthropic.com/872c653b2d0501d6ab44cf87f43e1dc4853e4d37.pdf}}
}

@misc{stixAIClosedDoors2025a,
  author       = {Stix, Charlotte and Pistillo, Matteo and Sastry, Girish and Hobbhahn, Marius and Ortega, Alejandro and Balesni, Mikita and Hallensleben, Annika and Goldowsky-Dill, Nix and Sharkey, Lee},
  title        = {{{AI Behind Closed Doors}}: A {{Primer}} on {{The Governance}} of {{Internal Deployment}}},
  year         = {2025},
  howpublished = {\url{http://arxiv.org/abs/2504.12170}},
  note         = {Accessed 2025-11-25}
}

@article{meinke_frontier_models_capable_incontext_scheming_2025,
  title        = {Frontier Models are Capable of In-context Scheming},
  author       = {Meinke, Alexander and Schoen, Bronson and Scheurer, J{\'e}r{\'e}my and Balesni, Mikita and Shah, Rusheb and Hobbhahn, Marius},
  journal      = {arXiv preprint},
  volume       = {arXiv:2412.04984},
  year         = {2025},
  doi          = {10.48550/arXiv.2412.04984},
  url          = {https://arxiv.org/abs/2412.04984}
}

@misc{iia_global_technology_audit_guides_2024,
  author       = {{Institute of Internal Auditors}},
  title        = {Global Technology Audit Guides (GTAGs)},
  year         = {2024},
  url          = {https://www.theiia.org/en/standards/2024-standards/global-guidance/},
  howpublished = {\url{https://www.theiia.org/en/standards/2024-standards/global-guidance/}},
  note         = {Part of the Global Guidance under the 2024 Global Internal Audit Standards}
}

@misc{fca_equifax_final_notice_2023,
  author       = {{Financial Conduct Authority}},
  title        = {Final Notice: Equifax Limited},
  year         = {2023},
  url          = {https://www.fca.org.uk/publication/final-notices/equifax-limited-2023.pdf},
  howpublished = {\url{https://www.fca.org.uk/publication/final-notices/equifax-limited-2023.pdf}},
  note         = {UK Financial Conduct Authority Final Notice}
}

@misc{iia_global_view_internal_audit_2022,
  author       = {{Institute of Internal Auditors}},
  title        = {Global View of Internal Audit: 2022 Executive Summary},
  year         = {2022},
  url          = {https://www.theiia.org/en/content/research/foundation/2022/global-view/},
  howpublished = {\url{https://www.theiia.org/en/content/research/foundation/2022/global-view/}},
  note         = {Institute of Internal Auditors Research Foundation}
}

@techreport{bcbs_governance_framework_d383_2016,
  author       = {{Basel Committee on Banking Supervision}},
  title        = {Corporate Governance Principles for Banks},
  institution  = {{Bank for International Settlements}},
  year         = {2016},
  url          = {https://www.bis.org/bcbs/publ/d383.pdf},
  howpublished = {\url{https://www.bis.org/bcbs/publ/d383.pdf}},
  note         = {BCBS Publication No. 383, Bank for International Settlements}
}

@misc{anthropic_sabotage_risk_report_internal_stress_testing_2025,
  author       = {{Anthropic}},
  title        = {Sabotage Risk Report: Pilot Risk Report -- Internal Stress Testing Team Review},
  year         = {2025},
  url          = {https://alignment.anthropic.com/2025/sabotage-risk-report/2025_pilot_risk_report_internal_stress_testing_team_review.pdf},
  howpublished = {\url{https://alignment.anthropic.com/2025/sabotage-risk-report/2025_pilot_risk_report_internal_stress_testing_team_review.pdf}},
  note         = {Internal stress testing team review}
}

@misc{monzo_annual_report_2018,
  author       = {{Monzo Bank Limited}},
  title        = {Annual Report 2018},
  year         = {2018},
  url          = {https://monzo.com/static/docs/annual-report-2018.pdf},
  howpublished = {\url{https://monzo.com/static/docs/annual-report-2018.pdf}}
}

@misc{monzo_annual_report_2021,
  author       = {{Monzo Bank Limited}},
  title        = {Annual Report 2021},
  year         = {2021},
  url          = {https://monzo.com/static/docs/monzo-annual-report-2021.pdf},
  howpublished = {\url{https://monzo.com/static/docs/monzo-annual-report-2021.pdf}}
}

@article{clymer_safety_cases_2024,
  title        = {Safety Cases: How to Justify the Safety of Advanced AI Systems},
  author       = {Clymer, Joshua and Gabrieli, Nick and Krueger, David and Larsen, Thomas},
  journal      = {arXiv preprint},
  volume       = {arXiv:2403.10462},
  year         = {2024},
  url          = {https://arxiv.org/abs/2403.10462},
  doi          = {10.48550/arXiv.2403.10462}
}

@misc{monzo_annual_report_2022,
  author       = {{Monzo Bank Limited}},
  title        = {Annual Report 2022},
  year         = {2022},
  url          = {https://monzo.com/static/docs/monzo-annual-report-2022.pdf},
  howpublished = {\url{https://monzo.com/static/docs/monzo-annual-report-2022.pdf}}
}

@misc{monzo_annual_report_2024,
  author       = {{Monzo Bank Limited}},
  title        = {Annual Report 2024},
  year         = {2024},
  url          = {https://monzo.com/annual-report/2024},
  howpublished = {\url{https://monzo.com/annual-report/2024}},
  note         = {Online annual report}
}

@misc{house_oversight_equifax_report_2018,
  author       = {{U.S. House Committee on Oversight and Government Reform}},
  title        = {The Equifax Data Breach:  Protecting Consumer Information in the Digital Age},
  year         = {2018},
  url          = {https://oversight.house.gov/wp-content/uploads/2018/12/Equifax-Report.pdf},
  howpublished = {\url{https://oversight.house.gov/wp-content/uploads/2018/12/Equifax-Report.pdf}},
  note         = {U.S. House of Representatives, Committee on Oversight and Government Reform}
}

@book{isaca_cobit_2019_framework,
  author       = {{ISACA}},
  title        = {COBIT 2019 Framework: Governance and Management Objectives},
  year         = {2019},
  publisher    = {ISACA},
  url          = {https://www.isaca.org/resources/cobit},
  howpublished = {\url{https://www.isaca.org/resources/cobit}}
}

@misc{emporia_types_internal_audits_2025,
  title        = {Types of Internal Audits},
  author       = {{Emporia State University}},
  year         = {2025},
  url          = {https://www.emporia.edu/internal-audit/types-internal-audits/},
  howpublished = {\url{https://www.emporia.edu/internal-audit/types-internal-audits/}}
}

@misc{anthropic_trust_portal_2025,
  title        = {Anthropic Trust Portal},
  author       = {{Anthropic}},
  year         = {2025},
  url          = {https://trust.anthropic.com/},
  howpublished = {\url{https://trust.anthropic.com/}}
}

@article{williamsWinningDefiningContest2025,
  author       = {Williams, Brandon Kirk},
  title        = {Winning the {{Defining Contest}}: {{The US-China Artificial Intelligence Race}}},
  year         = {2025},
  volume       = {48},
  number       = {2},
  pages        = {153--171},
  doi          = {10.1080/0163660X.2025.2517501},
  url          = {https://www.tandfonline.com/doi/full/10.1080/0163660X.2025.2517501},
  issn         = {0163-660X, 1530-9177},
  journaltitle = {The Washington Quarterly},
  shortjournal = {The Washington Quarterly},
  shorttitle   = {Winning the {{Defining Contest}}}
}

@misc{zvenyhorodskyiAIActionPlans2025,
  author       = {Zvenyhorodskyi, Pavlo, Scott Singer},
  title        = {The {{AI Action Plans}}: {{How Similar}} Are the {{U}}.{{S}}. and {{Chinese Playbooks}}?},
  year         = {2025},
  howpublished = {\url{https://www.justsecurity.org/119509/us-chinese-ai-playbooks/}},
  note         = {Accessed 2025-11-26}
}

@techreport{institute_of_internal_auditors_global_internal_audit_standard_2024,
  title = {Global internal audit standards},
  author = {{Institute of Internal Auditors}},
  year = {2024},
  url = {https://www.theiia.org/globalassets/site/standards/globalinternalauditstandards_2024january9.pdf},
  howpublished = {\url{https://www.theiia.org/globalassets/site/standards/globalinternalauditstandards_2024january9.pdf}}
}

@misc{homewood_third_party_compliance_reviews_2025,
  title = {Third-party compliance reviews for frontier AI safety frameworks},
  author = {Homewood, A. and Williams, S. and Dreksler, N. and Lidiard, J. and Murray, M. and Heim, L. and Ziosi, M. and Ó hÉigeartaigh, S. and Chen, M. and Wei, K. and Winter, C. and Brundage, M. and Garfinkel, B. and Schuett, J.},
  year = {2025},
  url = {https://doi.org/10.48550/arXiv.2505.01643},
  doi = {10.48550/arXiv.2505.01643},
  howpublished = {\url{https://doi.org/10.48550/arXiv.2505.01643}},
  note = {arXiv preprint}
}

@article{schuett_frontier_ai_developers_need_an_2024,
  title = {frontier AI developers need an internal audit function},
  author = {Schuett, J},
  year = {2024},
  url = {https://doi.org/10.1111/risa.17665},
  doi = {10.1111/risa.17665},
  howpublished = {\url{https://doi.org/10.1111/risa.17665}},
  note = {Risk Analysis. https://doi.org/10.1111/risa.17665}
}

@misc{european_commission_code_of_practice_final_version_2025,
  title = {Code of Practice: Final version – Commitments by Providers of General-Purpose AI Models with Systemic Risk, Commitment 8: Systemic Risk Responsibility Allocation},
  author = {{European Commission}},
  year = {2025},
  url = {https://digital-strategy.ec.europa.eu/en/policies/contents-code-gpai}
}

@misc{chartered_institute_of_internal_auditors_what_is_internal_audit_charter_2022,
  title = {What is internal audit? Chartered IIA},
  author = {{Chartered Institute of Internal Auditors.}},
  year = {2022},
  url = {https://charterediia.org/content-hub/blogs/what-is-internal-audit/},
  howpublished = {\url{https://charterediia.org/content-hub/blogs/what-is-internal-audit/}},
  note = {Accessed September 15, 2025},
  note = {Accessed September 15, 2025. https://charterediia.org/content-hub/blogs/what-is-internal-audit/}
}

@misc{the_institute_of_internal_auditors_staffing_considerations_for_in_2018,
  title = {Staffing Considerations for Internal Audit Activity},
  author = {{The Institute of Internal Auditors}},
  year = {2018},
  url = {https://www.theiia.org/globalassets/documents/resources/staffing-considerations-for-internal-audit-activity-may-2018/staffing-considerations-for-internal-audit-activity.pdf},
  howpublished = {\url{https://www.theiia.org/globalassets/documents/resources/staffing-considerations-for-internal-audit-activity-may-2018/staffing-considerations-for-internal-audit-activity.pdf}}
}

@misc{giattino_samborska_training_compute_2025,
  author       = {Giattino, Charlie and Samborska, Veronika},
  title        = {Since 2010, the Training Computation of Notable AI Systems Has Doubled Every Six Months},
  year         = {2025},
  month        = jan,
  day          = {21},
  url          = {https://ourworldindata.org/data-insights/since-2010-the-training-computation-of-notable-ai-systems-has-doubled-every-six-months},
  howpublished = {\url{https://ourworldindata.org/data-insights/since-2010-the-training-computation-of-notable-ai-systems-has-doubled-every-six-months}},
  note         = {Our World in Data. Accessed 30 November 2025}
}

@misc{metr_evaluating_frontier_ai_r_d_cap_2024,
  title = {Evaluating frontier AI R\&D capabilities of language model agents against human experts},
  author = {{METR}},
  year = {2024},
  url = {https://metr.org/},
  howpublished = {\url{https://metr.org/}}
}

@techreport{national_institute_of_standards_framework_for_improving_critic_2018,
  title = {Framework for Improving Critical Infrastructure Cybersecurity, Version 1},
  author = {{National Institute of Standards and Technology}},
  year = {2018},
  url = {https://doi.org/10.6028/NIST.CSWP.04162018},
  doi = {10.6028/NIST.CSWP.04162018},
  howpublished = {\url{https://doi.org/10.6028/NIST.CSWP.04162018}},
  note = {1. https://doi.org/10.6028/NIST.CSWP.04162018}
}

@misc{openai_gpt_4_system_card_2023,
  title = {GPT-4 System Card},
  author = {{OpenAI}},
  year = {2023},
  url = {https://cdn.openai.com/papers/gpt-4-system-card.pdf},
  howpublished = {\url{https://cdn.openai.com/papers/gpt-4-system-card.pdf}}
}

@misc{perez_red_teaming_language_models_wi_2022,
  title = {Red teaming language models with language models},
  author = {Perez, E., et al},
  year = {2022},
  url = {https://arxiv.org/abs/2202.03286},
  howpublished = {\url{https://arxiv.org/abs/2202.03286}},
  note = {arXiv preprint}
}

@misc{wei_jailbroken_how_does_llm_safety_2024,
  title = {Jailbroken: How does LLM safety training fail? Advances in Neural Information Processing Systems, 36},
  author = {Wei, A. and Haghtalab, N. and Steinhardt, J.},
  year = {2024},
  url = {https://arxiv.org/abs/2307.02483},
  howpublished = {\url{https://arxiv.org/abs/2307.02483}},
  note = {arXiv preprint}
}

@misc{anthropic_anthropic_s_responsible_scalin_2023,
  title = {Anthropic's Responsible Scaling Policy},
  author = {{Anthropic}},
  year = {2023},
  url = {https://www.anthropic.com/news/anthropics-responsible-scaling-policy},
  howpublished = {\url{https://www.anthropic.com/news/anthropics-responsible-scaling-policy}}
}
\clearpage

\appendix

\section{Risk-based internal audit model}\label{appendix:riskbasedaudit}

Risk-based internal auditing (RBIA) is the methodology by which internal audit functions align assurance activities with an organization's most significant risks. This appendix provides an overview of the risk-based audit process for readers unfamiliar with this aspect of internal audit. All guidance in this section is based on the guidance in the Institute of Internal Auditors' Global Internal Audit Standards (Standard 9.1, 2024) and provides illustrative examples, rather than complete or factually accurate ones.

The core principle of RBIA is resource allocation based on risk significance: audit effort should concentrate where residual risk is highest after considering inherent risk and control effectiveness (Griffiths, 2005). This contrasts with cyclical auditing (covering all areas on rotation) or compliance-driven auditing (responding to regulatory requirements), though both may inform risk-based prioritization.

\subsection{Stages of risk-based audit planning}

Risk-based audit planning follows a structured process from risk identification through to audit execution and reporting. The IIA's Practice Guide on Developing the Internal Audit Strategic Plan (2012) and Implementation Guide 2010 (2017) establish the foundational methodology.

\subsubsection{Stage 1: Understand the organization and its context}

The first stage requires an internal audit to develop a comprehensive understanding of the objectives, strategy, operating environment, and risk appetite of the organization.

\textbf{Information access required:} Primarily structural, such as strategic plans, board materials, published safety frameworks (RSP or equivalent), regulatory filings, public commitments.

\subsubsection{Stage 2: Identify and assess risks}

The second stage involves identifying and assessing risks, typically ranked by likelihood and impact. Where prior audits have been done, report ratings can inform current estimates of residual risk. This risk register forms the basis of risk-based audit planning by driving the prioritization of risks for review. The following illustrative register shows how a sample of frontier AI risks may be documented and assessed.

\begin{table}[H]
\centering
\caption{Illustrative Risk Register}
\small
\setlength{\tabcolsep}{8pt}
\begin{tabular}{@{}p{0.50\textwidth} p{0.12\textwidth} p{0.10\textwidth} p{0.15\textwidth}@{}}
\toprule
\textbf{Risk Description} & \textbf{Risk Rating} & \textbf{Trend} & \textbf{Last Assessed} \\
\midrule
External misuse of cyber-offensive capabilities enables catastrophic cyberattack & 15 & $\uparrow$ & Q4 2025 \\
\addlinespace[0.5em]
Model weights exfiltrated by state actor enabling unrestricted model access & 15 & $\rightarrow$ & Q3 2025 \\
\addlinespace[0.5em]
Internal misuse of unreleased model for unauthorized purposes & 8 & $\rightarrow$ & - \\
\addlinespace[0.5em]
Dangerous capability evaluation fails to identify material capability & 10 & $\uparrow$ & - \\
\addlinespace[0.5em]
CBRN capabilities enable weapons development assistance & 15 & $\uparrow$ & - \\
\bottomrule
\end{tabular}
\end{table}

\textbf{Information access required:} Primarily operational, such as enterprise risk management outputs, incident history, near-miss data, capability evaluation results, security assessments, regulatory guidance.

\subsubsection{Stage 3: Construct the audit universe}

The audit universe is a comprehensive inventory of all auditable units within the organization. Each auditable unit represents a discrete area that could be subject to an internal audit engagement, for example an area, entity, operation, function, process, or system (IIA, 2024a).

The following is an illustrative example of how a simplified version of an audit universe could be constructed for the risk: 'External misuse of cyber-offensive capabilities enables catastrophic cyberattack'.

Map risk pathways to set out how a risk could materialize, helping to identify the auditable units that sit along this path.

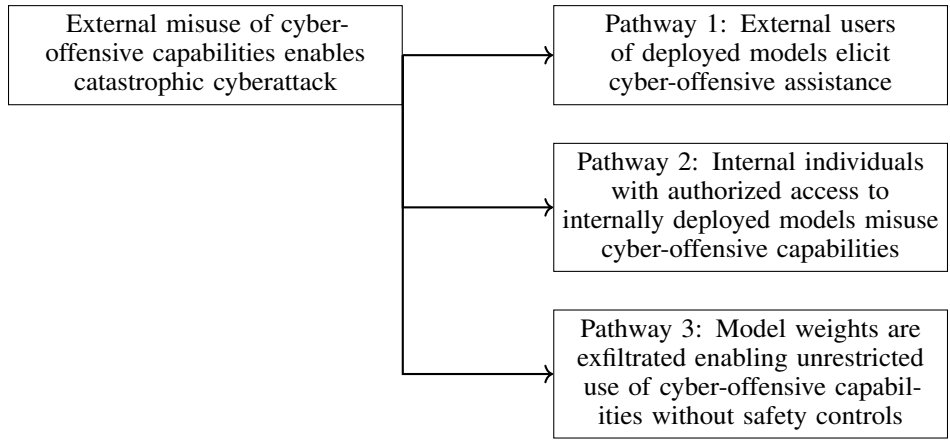
\begin{figure}[H]
\centering
\begin{tikzpicture}[
    node distance=0.5cm,
    box/.style={rectangle, draw, text width=5cm, align=center, minimum height=1cm},
    arrow/.style={->, thick}
]
    \node[box] (main) {External misuse of cyber-offensive capabilities enables catastrophic cyberattack};
    \node[box, right=2cm of main] (p1) {Pathway 1: External users of deployed models elicit cyber-offensive assistance};
    \node[box, below=of p1] (p2) {Pathway 2: Internal individuals with authorized access to internally deployed models misuse cyber-offensive capabilities};
    \node[box, below=of p2] (p3) {Pathway 3: Model weights are exfiltrated enabling unrestricted use of cyber-offensive capabilities without safety controls};
    
    \draw[arrow] (main.east) -- (p1.west);
    \draw[arrow] (main.east) |- (p2.west);
    \draw[arrow] (main.east) |- (p3.west);
\end{tikzpicture}
\caption{Risk pathways for external misuse of cyber-offensive capabilities}
\end{figure}

A pathway can then be broken down into a risk chain to identify the steps which lead to harm and the key control points which act as layers of defense against the risk materializing. For example Pathway 1 could have the following risk chain:

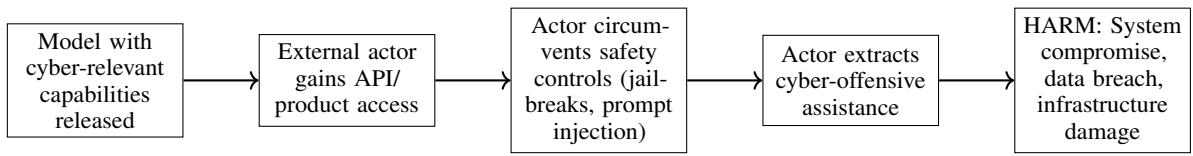
\begin{figure}[H]
\centering
\begin{tikzpicture}[
    node distance=0.3cm and 1cm,
    box/.style={rectangle, draw, text width=2.1cm, align=center, minimum height=1.2cm, font=\small},
    arrow/.style={->, thick}
]
    \node[box] (n1) {Model with cyber-relevant capabilities released};
    \node[box, right=of n1] (n2) {External actor gains API/ product access};
    \node[box, right=of n2] (n3) {Actor circumvents safety controls (jailbreaks, prompt injection)};
    \node[box, right=of n3] (n4) {Actor extracts cyber-offensive assistance};
    \node[box, right=of n4] (n5) {HARM: System compromise, data breach, infrastructure damage};
    
    \draw[arrow] (n1) -- (n2);
    \draw[arrow] (n2) -- (n3);
    \draw[arrow] (n3) -- (n4);
    \draw[arrow] (n4) -- (n5);
\end{tikzpicture}
\caption{Risk chain for Pathway 1}
\end{figure}

Once the steps in the risk chain and key control points have been identified, these act as candidates for areas for internal audit to review ('auditable units'). Each may be a control for a single or multiple pathways across one or more risks, for example incident response capability.

\begin{table}[H]
\centering
\caption{Example auditable units}
\small
\setlength{\tabcolsep}{6pt}
\begin{tabular}{@{}p{0.22\textwidth} p{0.13\textwidth} p{0.20\textwidth} p{0.38\textwidth}@{}}
\toprule
\textbf{Auditable Unit} & \textbf{Audit Level} & \textbf{Control Point in Risk Pathway} & \textbf{Assurance Provided} \\
\midrule
Dangerous capability evaluation (cyber) & Model & Capability identification & Evaluations accurately identify cyber-offensive capabilities before deployment \\
\addlinespace[0.8em]
Safety classifier design and effectiveness & Model & Output restriction & Classifiers reliably block cyber-offensive outputs under adversarial conditions \\
\addlinespace[0.8em]
Post-deployment monitoring & System & Detection & Monitoring detects cyber-offensive misuse patterns in production \\
\addlinespace[0.8em]
Incident response capability & System & Response & Organization can contain and remediate confirmed cyber misuse effectively \\
\addlinespace[0.8em]
Deployment decision governance & Governance & Decision-rights & Deployment decisions appropriately weigh cyber capability risks \\
\bottomrule
\end{tabular}
\end{table}

\textbf{Access to information required:} A mix of organizational, procedural, operational and technical information would be required to build an accurate picture of the teams and systems who contribute to the controls in place to manage a risk.

\subsubsection{Stage 4: Develop and maintain the audit plan}

The fourth stage translates these outputs into an audit plan. The audit plan specifies which auditable units will be covered in the planning period, with indicative timing, resource allocation, and skill requirements. The audit plan is generally created by the Chief Audit Executive (CAE) and approved by the board, or (typically risk) committee they have delegated responsibility to. The audit plan should reflect the areas of the organization most important for internal audit to provide assurance over and should be monitored and updated if there are changes to organization's business, risks, operations, programs, systems, and controls (IIA, Standard 9.1).

Below is an illustrative example of a sample of an audit plan:

\begin{table}[H]
\centering
\caption{Illustrative audit plan sample}
\footnotesize
\setlength{\tabcolsep}{5pt}
\begin{tabular}{@{}p{0.08\textwidth} p{0.16\textwidth} p{0.10\textwidth} p{0.08\textwidth} p{0.30\textwidth} p{0.18\textwidth}@{}}
\toprule
\textbf{Timing} & \textbf{Auditable Unit} & \textbf{Audit Level} & \textbf{Days} & \textbf{Information Access Required} & \textbf{Rationale} \\
\midrule
Q1 & Dangerous capability evaluation (cyber) & Model & 100 & \textit{Technical:} Evaluation benchmarks, model performance data, uplift measurements. \textit{Procedural:} Evaluation methodology, trigger criteria. \textit{Operational:} Completed evaluations, evaluator interviews. & Highest residual risk; increasing trend; foundational assurance \\
\addlinespace[0.8em]
Q2 & Safety classifier effectiveness & Model & 100 & \textit{Technical:} Classifier architecture, training data, performance metrics, adversarial test results. \textit{Procedural:} Development procedures, monitoring procedures. \textit{Operational:} Production metrics, evasion incident records. & High residual risk; increasing trend; critical control for external misuse \\
\addlinespace[0.8em]
Q3 & Incident response capability & System & 80 & \textit{Structural:} IR policy, team charter. \textit{Procedural:} IR procedures, playbooks. \textit{Operational:} Incident logs, tabletop exercise results, IR team interviews. \textit{Technical:} Containment mechanisms, rollback capabilities. & High residual risk; untested capability \\
\bottomrule
\end{tabular}
\end{table}

\textbf{Access to information required:} A mix of organizational, procedural, operational and technical information to track changes to factors impacting the organization's risk profile, for example strategic plan updates, governance changes, revised policies and procedures, incident and near-miss data, audit findings, control performance metrics, new system deployments and infrastructure changes.

\subsubsection{Stage 5: Execute audits and report}

Once the audit plan has been agreed with the board, the individual audit engagements within the plan can be executed. Each audit engagement involves detailed planning, fieldwork, and reporting. During fieldwork, internal auditors identify the key controls within an area which are being relied on to manage the risk and will evaluate how well these are both designed ('design effectiveness') and operating in practice ('operational effectiveness').

An audit report is expected to provide 'reasonable assurance', based on having sufficient and appropriate evidence, intended to address the risk of expressing inaccurate assurance statements (ISACA, ICAEW, n.d.). Some internal audit standards (ICAEW, n.d.; IIA, 2024) allow for a low level of assurance, 'limited assurance' to be provided, based upon the nature, timing, and extent of procedures performed, including limited access to evidence.

Findings are reported to management and the board, with follow-up on remediation actions. Here is an illustrative example of how a finding may be reported in an audit report:

\begin{table}[H]
\centering
\caption{Illustrative audit finding}
\footnotesize
\setlength{\tabcolsep}{5pt}
\begin{tabular}{@{}p{0.10\textwidth} p{0.21\textwidth} p{0.14\textwidth} p{0.14\textwidth} p{0.18\textwidth} p{0.13\textwidth}@{}}
\toprule
\textbf{Risk Rating} & \textbf{Finding} & \textbf{Impact} & \textbf{Recommendation} & \textbf{Management Response} & \textbf{Owner} \\
\midrule
Medium & Incident logs do not retain sufficient conversation context for effective post-incident forensic analysis; key data was purged under standard retention before investigation. & Limits root-cause analysis and control improvement; increases likelihood of repeat exploitation. & Extend retention for flagged sessions (e.g. 90 days); introduce formal forensic hold on incident initiation. & Agreed. Enhanced retention and forensic hold to be added to IR playbook, subject to legal review. Due: Q1 2026 & Chief Information Security Officer \\
\bottomrule
\end{tabular}
\end{table}

\textbf{Access to information required:} Audit dependent for fieldwork and control testing, could span organizational, procedural, operational and technical information depending on the audit scope. Evidence will also be required to confirm that audit findings have been closed.

\newpage

\section{Planning frequency decisions for internal audit} \label{appendix:auditplanningfrequencies}

While the audit plan itself does not generate assurance, the frequency with which it is reviewed and updated determines whether internal audit resources remain directed toward the organization's highest-priority risks. If the plan is not revisited as the risk environment evolves, internal audit may continue providing assurance over areas that no longer warrant priority attention, while emerging risks remain unexamined. Planning frequency therefore directly impacts the relevance of assurance provided by the Internal Audit Function (IAF). It should be noted that costs of more frequent audit plan updates are generally much lower than the costs of more frequent audit execution, as planning activities primarily require Chief Audit Executive (CAE) time rather than full audit team deployment.

An examination of audit planning practices reveals three major approaches to the frequency with which audit plans are drawn up and revised: annual audit plans, annual audit plans with periodic updates, and rolling risk assessment and continuous plan updates. The relevance of each to IAF within frontier AI companies is discussed below.

\begin{table}[H]
\centering
\caption{Internal Audit Planning Frequency Options}
\small
\setlength{\tabcolsep}{6pt}
\begin{tabular}{@{}p{0.18\textwidth} p{0.26\textwidth} p{0.26\textwidth} p{0.23\textwidth}@{}}
\toprule
\textbf{Approach} & \textbf{Advantages} & \textbf{Disadvantages} & \textbf{Recommendation} \\
\midrule
Annual audit plan & Meets IIA minimum standards. & Cannot respond to emerging risks between annual reviews; insufficient given rapid pace of frontier AI development \& low cost of more frequent audit planning. & Potential baseline, from which to monitor how much risk ratings underpinning audit plan change over time. \\
\addlinespace[0.8em]
Annual audit plan with periodic updates & Preserves resource allocation and scheduling benefits of annual framework; enables responsive adjustment to emerging risks. & Less adaptive than continuous monitoring. & Well-suited to frontier AI companies establishing an IAF. Minimum appropriate frequency. \\
\addlinespace[0.8em]
Rolling risk assessment and continuous plan updates, & Fastest possible response to emerging risks; aligns with continuous auditing approach. & More resource intensive; requires sophisticated risk intelligence capabilities and monitoring infrastructure. & Ambitious but potentially targets most critical areas for assurance for frontier AI companies. \\
\bottomrule
\end{tabular}
\end{table}

\subsection{Annual audit plan}

IIA standards require risk assessments and audit plans to be developed at least annually (IIA, 2024), establishing audit plans drawn up once a year and executed with no subsequent amendments as the baseline frequency for audit planning. Such annual cycles align with corporate governance rhythms in most organizations and may suffice for smaller organizations or lower-risk sectors. However, annual planning is widely regarded as a floor rather than a ceiling in dynamic industries, particularly those experiencing rapid change (Khayal, 2022; IIA). However, many CAEs in fast-moving industries update plans periodically or on a rolling basis (IIA, 2020); both these options are discussed below.

\subsection{Annual audit plan with periodic updates}

This approach involves drafting an audit plan once a year and revisiting and possibly modifying that plan at predetermined times within the year, such as every quarter (IIA, 2020), or even monthly if the organization's environment is dynamic (IIA, 2024). During these reviews, the CAE and audit committee can evaluate whether planned audits remain appropriate given recent developments, whether emerging risks require additional ad hoc audits (see below) and whether resource allocation within the IAF should be adjusted.

Annual planning with periodic updates preserves the resource allocation and scheduling benefits of an annual framework while enabling responsive adjustment to emerging risks. As such, this approach appears well-suited to frontier AI developers seeking to establish an IAF.

\subsection{Rolling risk assessment and continuous plan updates}

The most adaptive approach treats audit planning as a continuous process rather than a periodic event. Under this model, the IAF maintains ongoing monitoring of the risk environment and updates audit priorities dynamically as risks emerge or evolve (IIA, 2015). Rather than waiting for a pre-arranged review to adjust the audit plan, adjustments can be made in near real-time. This approach enables the fastest possible response to emerging risks, and aligns well with the "continuous auditing" approach to execution frequency, as explored below. However, such an approach is necessarily more resource intensive and requires more sophisticated risk intelligence capabilities. As such, this approach is most feasible for well-resourced IAFs with strong risk monitoring infrastructure.

Rolling risk assessment therefore represents a more ambitious approach to audit planning, suited to frontier AI developers with mature IAF capabilities and robust risk monitoring infrastructure.

\newpage

\section{Options for Managing Auditor Access to Sensitive Information} \label{appendix:accesscontrols}

This appendix outlines options for enabling auditor access to sensitive information while managing associated security risks.

\subsection{Tiered Access Model}

Assurance activities do not require all evaluators to access all forms of sensitive information. A tiered access model, grounded in least-privilege principles, grants individuals only the minimum access necessary to perform their defined assurance role. This approach draws on precedents such as the U.S. Federal Reserve's Confidential Supervisory Information framework, which restricts access based on role, purpose, and sensitivity (12 C.F.R. § 261.2, 2024).

\begin{table}[H]
\centering
\caption{Tiered access model for auditors}
\small
\setlength{\tabcolsep}{6pt}
\begin{tabular}{@{}p{0.18\textwidth} p{0.26\textwidth} p{0.24\textwidth} p{0.24\textwidth}@{}}
\toprule
\textbf{Tier} & \textbf{Scope} & \textbf{Access modality} & \textbf{Assurance focus} \\
\midrule
Tier 1: Generalist Auditors & Governance documents, policies, HR and finance records & Read-only access through standard corporate systems & Governance and process-level controls \\
\addlinespace[0.8em]
Tier 2: Technical Evaluators & Model weights, red-teaming logs, dangerous capability evaluations, security incident forensics & Clean room environments with no external network access or data export & System-level and model-level risk adequacy \\
\bottomrule
\end{tabular}
\end{table}

\subsection{Clean Room Environment}

For highly sensitive technical information, review can occur within a secure enclave designed to permit examination while preventing exfiltration. This model draws on established precedents including the FTC's clean team protocols for antitrust data rooms (FTC, 2018), algorithmic IP protections in high-frequency trading (FINRA, 2015), and Sensitive Compartmented Information Facilities (SCIFs) for classified information (ODNI, 2010).

\begin{table}[H]
\centering
\caption{Clean room control measures}
\small
\setlength{\tabcolsep}{8pt}
\begin{tabular}{@{}p{0.20\textwidth} p{0.75\textwidth}@{}}
\toprule
\textbf{Control type} & \textbf{Measures} \\
\midrule
Technical & Virtual Desktop Infrastructure (VDI) with network segregation; allow listed access to specific model APIs only; disabled clipboard, external drives, and screen capture \\
\addlinespace[0.8em]
Physical & For highest-sensitivity materials (e.g., weight encryption keys, novel vulnerability details), physical presence in SCIF at developer headquarters \\
\addlinespace[0.8em]
Monitoring & All evaluator activity logged; logs enable detection of IP misuse and provide auditable record of access scope; logging excludes credential capture \\
\addlinespace[0.8em]
Evidence integrity & Data access, modification, and logging conform to recognized chain-of-custody standards (ISO, 2017) \\
\bottomrule
\end{tabular}
\end{table}

\subsection{Time-Bounded and Rate-Limited Access}

Evaluators may be granted access that is both time-bounded and rate-limited, permitting inspection of sensitive artifacts (e.g., hidden reasoning traces, restricted evaluation endpoints) for a defined period and at a controlled query volume. This reduces the risk of large-scale extraction while enabling meaningful assurance. Apollo Research's rate-limited access to internal chain-of-thought traces for OpenAI o3 and o4-mini during their scheming evaluation provides a practical precedent (OpenAI, 2025).

\begin{table}[H]
\centering
\caption{Recommended access models by information type}
\small
\setlength{\tabcolsep}{8pt}
\begin{tabular}{@{}p{0.18\textwidth} p{0.20\textwidth} p{0.55\textwidth}@{}}
\toprule
\textbf{Information type} & \textbf{Typical sensitivity} & \textbf{Recommended access model} \\
\midrule
Structural & Low--Medium & Standard corporate system access \\
\addlinespace[0.5em]
Procedural & Medium & Standard access with NDA; restricted documentation handling \\
\addlinespace[0.5em]
Operational & Medium--High & Controlled access; interview protocols; communication review under supervision \\
\addlinespace[0.5em]
Technical & High & Clean room environment; tiered access; time-bounded and rate-limited; session monitoring \\
\bottomrule
\end{tabular}
\end{table}

\clearpage

\begin{landscape}
\section{Illustrative evidence required for reasonable assurance by lifecycle} \label{appendix:evidencebylifecycle}
\begin{table}[H]
\centering
\caption{Evidence requirements across AI lifecycle phases}
\label{tab:evidence_lifecycle}
\tiny
\setlength{\tabcolsep}{4pt}

\begin{adjustbox}{max width=\linewidth}
\begin{tabular}{@{}p{0.12\linewidth} p{0.14\linewidth} p{0.10\linewidth} p{0.32\linewidth} p{0.32\linewidth}@{}}
\toprule
\textbf{Phase} & \textbf{Example Evidence} & \textbf{Information Type} & \textbf{Description} & \textbf{Example Control Test} \\
\midrule

\multirow{3}{*}{1. Governance \& Strategy} 
& AI Governance Charter \& Safety Framework & Structural & Documented framework defining acceptable development and deployment thresholds, including capability limits and required safeguards at each level. & Identify documented instances where the framework halted or materially altered a development or deployment decision. \\
\addlinespace[0.5em]
& Board \& Risk Committee Minutes & Structural & Unredacted records of board and committee discussions on AI risk, including deliberations, challenges raised, and decisions taken. & Determine whether any deployment proceeded despite documented unresolved safety concerns or dissenting views. \\
\addlinespace[0.5em]
& Risk Appetite Statement & Structural & Quantified limits on acceptable risk exposure across risk categories (e.g., zero tolerance for CBRN assistance, defined uplift thresholds for cyber capabilities). & Compare actual deployment decisions against stated quantitative thresholds to identify any breaches or exceptions. \\
\addlinespace[1em]

\multirow{3}{*}{2. Data Governance}
& Data Lineage \& Provenance Records & Procedural & System logs documenting the origin, transformation, and chain of custody for training data throughout the pipeline. & Trace specific model outputs of concern back through the data pipeline to identify source content and validate filtering controls. \\
\addlinespace[0.5em]
& Data Privacy Impact Assessments (DPIAs) & Procedural & Documented legal assessments of personal-data usage, processing lawfulness, and risk mitigation measures. & Verify that data subject requests (e.g., erasure under GDPR Article 17) were processed within required timeframes. \\
\addlinespace[0.5em]
& Filtering \& Cleaning Logs & Procedural & Records of content filtered, flagged, or discarded during data curation, including filter rules applied and volumes affected. & Confirm that filtering mechanisms detected and removed known categories of harmful content from training data. \\
\addlinespace[1em]

\multirow{3}{*}{3. Model Development}
& Unreleased Model Weights & Technical & Snapshots of trained model parameters at defined checkpoints throughout development. & Verify cryptographic integrity between the safety-evaluated model version and the version deployed to production. \\
\addlinespace[0.5em]
& MLOps Change Logs & Technical & Version-control history for training code, configuration files, and pipeline modifications. & Review commit history for unauthorized or undocumented changes that could introduce vulnerabilities or backdoors. \\
\addlinespace[0.5em]
& Bias Testing Reports & Technical & Intermediate evaluation results assessing model performance across demographic groups and sensitive attributes. & Identify early indicators of fairness issues and verify that flagged concerns were addressed before deployment. \\
\addlinespace[1em]

\multirow{2}{*}{4. Pre-Deployment Evaluation}
& Unredacted Red Teaming Reports & Operational & Raw logs and detailed findings from adversarial testing, including specific attack vectors attempted and model responses. & Review attack content, assess severity of successful elicitations, and verify that remediations address root causes rather than specific prompts. \\
\addlinespace[0.5em]
& Dangerous Capability Evaluations & Technical & Benchmark scores and detailed results for biological, chemical, cyber, radiological, and persuasion capability assessments. & Confirm that evaluators employed rigorous elicitation techniques and that results accurately reflect model capabilities. \\
\addlinespace[1em]

\multirow{2}{*}{5. Post-Deployment Monitoring}
& AI Audit Trail \& SIEM Logs & Technical & Logs of API access patterns, query content, infrastructure activity, and security events from production systems. & Investigate near-miss events where attacks were detected and blocked to assess attacker behavior and control effectiveness. \\
\addlinespace[0.5em]
& Incident Reports & Operational & Documentation of confirmed misuse incidents, safety failures, or control bypasses, including root-cause analysis and remediation actions. & Verify that incidents triggered appropriate responses including model updates, control enhancements, or governance-policy revisions. \\
\bottomrule
\end{tabular}
\end{adjustbox}

\end{table}

\end{landscape}
\clearpage

\end{document}